\documentclass[prd,aps,floatfix,nofootinbib,11 pt]{revtex4}
\usepackage{amsmath,graphicx,epsfig}

\begin{document}

\title{ Rare decays of $\Lambda_b \to \Lambda + \gamma$
and $\Lambda_b \to \Lambda + l^{+} l^{-}$ in the light-cone sum
rules}

\author{Yu-Ming Wang$^{1}$} \author{Ying Li$^{2}$} \author{Cai-Dian L\"{u} $^{1}$}

\vspace*{1.0cm}

\affiliation{$^{1}$Institute of High Energy Physics, P.O. Box
918(4), Beijing 100049, China}

\affiliation{$^{2}$Physics Department, Yantai University, Yantai,
264005, China}

\vspace*{1.0cm}

\begin{abstract}

The weak decays of $\Lambda_b \to \Lambda + \gamma$ and $\Lambda_b
\to \Lambda + l^{+} l^{-}$ are investigated in the Standard Model
using light-cone sum rules approach.  The higher twist distribution
amplitudes of $\Lambda$ baryon to the leading conformal spin are
included in the sum rules for transition form factors. Our results
indicate that the higher twist distribution amplitudes almost have
no influences on the transition form factors retaining the heavy
quark spin symmetry, while such corrections can result in
significant impacts on the form factors breaking the heavy quark
spin symmetry.  Two phenomenological models (COZ and FZOZ) for the
wave function of $\Lambda$ baryon are also employed in the sum rules
for a comparison, which can give rise to the    form factors
approximately five times  larger than that in terms of conformal
expansion. Utilizing the form factors calculated in LCSR, the
physical observables like  decay rate, polarization asymmetry and
forward-backward asymmetry are analyzed for the decays of $\Lambda_b
\to \Lambda \gamma$, $\Lambda l^{+}l^{-}$.

\end{abstract}

 \maketitle


\section{Introduction}

 Generally,   new physics can be accessible through rare decays, where the
contributions from the Standard Model (SM) are suppressed enough.
Hence, such decays can provide an ideal platform to test the SM
precisely as well as to bound new physics parameters stringently.
Rare decays involving $b \to s$ flavor changing neutral current
(FCNC), which are forbidden at the tree level in the standard model,
can only be  induced by Glashow-Iliopoulos-Maiani  mechanism
\cite{Glashow:1970gm} via loop diagrams. The
Cabibbo-Kobayashi-Maskawa (CKM) matrix \cite{CKM 1, CKM 2} elements
can be determined quantitatively from $b \to s$ rare decays,
$B_s^0-\bar{B_s^0}$ mixing \cite{Ali review} etc., which will test
its unitarity under the requirement of the SM.

It is well known that the inclusive decays are relatively robust
theoretically, since the decay rate can be systematically and
reasonably approximated by the decay of a free $b$ quark into light
quarks, gluons and photons;  but the counterpart of experimental
measurements are quite difficult. On the contrary, the experimental
investigations of exclusive decays are comparably easier; while the
theoretical analysis of these modes encounter formidable
difficulties due to lack of good understanding of QCD at low
energy. 
As a matter of fact, there have been extensive studies on the
exclusive decays of $B \to M \, l^{+} l^{-}$ \cite{b to s in theory
16,b to s in theory 17,b to s in theory 18,b to s in theory 19,b to
s in theory 20,b to s in theory 21}, $B \to V(A) \, \gamma$ \cite{b
to s in theory 31,b to s in theory 32,b to s in theory 33,b to s in
theory 34,b to s in theory 35,b to s in theory 36,b to s in theory
37,b to s in theory 38,b to s in theory 39,b to s in theory 40,b to
s in theory 41}   in the literature with varying degrees of
theoretical rigor and emphasis. Unlike mesonic decays, the
investigations of FCNC $b \to s$ transition for bottom baryonic
decays $\Lambda_b \to \Lambda \gamma$ and $\Lambda_b \to \Lambda
l^{+} l^{-}$ are much behind because more degrees of freedom are
involved in the bound state of baryon system at the quark level. It
should be pointed out that such baryonic decays can offer the unique
ground to extract  the helicity structure of effective Hamiltonian
for FCNC $b \to s$ transition in the SM and beyond, which is lost in
the hadronization of meson case.  From the viewpoint of experiment,
the only drawback of bottom baryon decays is that the production
rate of $\Lambda_b$ baryon in $b$ quark hadronization is about four
times less than that of the $B$ meson.

The polarization asymmetry and forward-backward asymmetry in rare
decays are very sensitive to the new physics effects beyond the SM
\cite{c.q. geng 1,c.q. geng 2,c.q. geng 3,c.q. geng 4, Aliev 1,Aliev
2,Aliev 3,Aliev 4}. For instance, the polarization asymmetry of
$\Lambda$ baryon in $\Lambda_b \to \Lambda + l^{+} l^{-}$ decays is
dependent heavily on the right-handed current \cite{c.q. geng 2},
which is suppressed in the SM. However, before claiming the new
physics signal, one should have a serious and comprehensive scrutiny
of discrepancies between experimental data and theoretical
predictions within the SM to confirm  whether they are indeed due to
contributions from new physics or missing effects in the SM, such as
higher power and next-to-leading order corrections, soft gluon
effects and so on.

 Currently, there has been  some studies in the
literature on $\Lambda_b \to \Lambda$ transition form factors
ranging from phenomenological models including the pole model (PM)
\cite{Mannel}, covariant oscillator quark model
(COQM)\cite{Mohanta}, MIT bag model (BM)\cite{Cheng} and
non-relativistic quark model \cite{Cheng 2}  to QCD sum rule
approach (QCDSR) \cite{Huang} and also perturbative QCD (PQCD)
approach \cite{HLLW}. It can be observed that the available
theoretical predictions vary from each other and can differ even by
the orders of magnitude. For instance, the decay width of $\Lambda_b
\to \Lambda \gamma$ in PQCD approach is about one to two orders
smaller than that obtained within other frameworks.

It is suggested that the soft non-perturbative contribution to the
transition form factor can be calculated quantitatively in the
framework of light-cone sum rules (LCSR) approach \cite{LCSR 1, LCSR
2,LCSR 3,LCSR 4,LCSR 5}, which is  a fully relativistic approach and
well rooted in quantum field theory, in a systematic and almost
model-independent way. As a marriage of standard QCDSR technique
\cite{SVZ 1, SVZ 2,SVZ 3} and theory of hard exclusive process
\cite{hard exclusive process 1,hard exclusive process 2,hard
exclusive process 3,hard exclusive process 4,hard exclusive process
5,hard exclusive process 6,hard exclusive process 7,hard exclusive
process 8}, LCSR cure the problem of QCDSR applying to the large
momentum transfer by performing the operator product expansion (OPE)
in terms of twist of the revelent operators rather than their
dimension \cite{braun talk}. Therefore, the principal discrepancy
between QCDSR and LCSR consists in that non-perturbative vacuum
condensates representing the long-distance quark and gluon
interactions in the short-distance expansion are substituted by the
light cone distribution amplitudes (LCDAs) describing the
distribution of longitudinal momentum carried by the valence quarks
of hadronic bound system in the expansion of transverse-distance
between partons in the infinite momentum frame. Phenomenologically,
LCSR has been widely applied to investigate the transition of mesons
and baryons in the recent years, such as bottom meson decays
\cite{Ball:1997rj,Ball:1998tj,Khodjamirian:2000ds,Ball:2004ye,
Melic:2004ud,Wang:2007fs}, electromagnetic form factors of nucleon
\cite{form factor of nucleon, Lenz:2003tq, Braun:2006hz}, pion
electroproduction \cite{Braun:2006td}, semi-leptonic weak decays of
$\Lambda_b \to p l \nu_{l}$ \cite{m.q. huang lambda_b to proton},
$\Lambda_c \to \Lambda l \nu_{l}$ \cite{m.q. huang}, and can make
consistent predictions with experimental data successfully. The
generalization to the $\gamma^{\ast} N \to \Delta$ transition
\cite{Braun:2005be}, $\Sigma \to N$ transition \cite{Wang:2006yz},
radiative decays of decuplet to octet baryons \cite{Aliev:2004ju}
and meson-octet baryon couplings \cite{Aliev:2006xr} have also  been
worked out.

  LCDAs are the fundamental non-perturbative functions in the LCSR,
which describes the hadronic structure in rare parton configurations
with a fixed number of Fock components at small transverse
separation in the infinite momentum frame. There have been
continuous interests concentrating on the research of pre-asymptotic
corrections to the distribution amplitudes of hadrons in the
exclusive reactions over two decades, in an attempt to accommodate
the experimental data, such as the electromagnetic form factors of
pion and nucleon \cite{hard exclusive process 8}. In particular, the
LCDAs of mesons including higher twist  have been investigated
extensively   \cite{hard exclusive process 8, DAs of meson original
2, DAs of meson original 3,DAs of meson develop 1, DAs of meson
develop 2, DAs of meson develop 3,Ball:2006wn, Ball:2007zt}. In
contrast to the meticulous and intensive study of mesonic LCDAs, the
corresponding task for the baryonic counterpart received less
attention in the past due to the complexity of the inner structures
for baryon. Fortunately, an exciting attempt to study the LCDAs of
baryon systematically was carried out in \cite{DAs of nucleon 1} for
the first time, where all the transverse degrees of freedom are
integrated out and the higher twist distribution amplitudes
contributed by the light-ray operators involving ``minus''
components of the quark field operators were investigated in detail.
Subsequently, an alternative approach to classify the three-quark
light-cone amplitudes of proton was put forward  in  \cite{x.d.ji
1,x.d.ji 2, x.d.ji 3}, where the transverse degree of freedom  of
partons is retained and the minus component of field operators are
eliminated in favor of the plus and transverse ones with the help of
equations of motion. It is mentioned in Ref. \cite{necleion DAs
review} that both techniques are probably equivalent.  In the
present work, we   would like to follow the prescription presented
in \cite{DAs of nucleon 1} for LCSR approach to study the rare decay
of $\Lambda_b \to \Lambda + \gamma$ and $\Lambda_b \to \Lambda +
l^{+} l^{-}$, which can be also generalized  to the decays of
$\Lambda_b$ to heavier  $\Lambda$ baryons
\cite{Legger:2006cq,Hiller:2007ur}, once the distribution amplitudes
of corresponding baryons are available.

The organization of this paper is as following: In section II, we
collect the effective Hamiltonian responsible for FCNC transition of
$b \to s \gamma$ and $b \to s l^{+}l^{-}$. Parameterizations of
various hadronic matrix elements $\langle \Lambda|\bar{s}\Gamma_i
b|\Lambda_b \rangle$ with $\Gamma_i$ being all the possible Lorentz
structures are also presented here. Section III  contains the
derivation of LCSR for both $\Lambda_b \to \Lambda + \gamma$ and
$\Lambda_b \to \Lambda + l^{+} l^{-}$ decays up to the leading order
of $\alpha_s$ with the corrections from higher twists. We   find
that the relations between various transition form factors existing
in the heavy quark effective theory (HQET) and large energy
effective theory (LEET) are well respected within the framework of
LCSR approach. The numerical analysis of LCSR for transition form
factors are displayed in section IV, where the corrections from the
distribution amplitudes of higher twist and comparisons with the
results that derived in other methods are investigated at length. In
section V, we apply the available form factors to study the
invariant mass distribution of lepton pair, $\Lambda$ polarization
asymmetry and forward-backward asymmetry of $\Lambda_b \to \Lambda +
l^{+} l^{-}$ together with the decay width and $\Lambda$
polarization asymmetry of $\Lambda_b \to \Lambda \gamma$. We confirm
the finding in Ref. \cite{Huang,x.g. he} that the $\Lambda$
polarization asymmetry of $\Lambda_b \to \Lambda \gamma$ is
completely independent on the hadronic transition form factors,
which makes it a good quantity to study the chiral structure of $b
\to s$ effective Hamiltonian and probe the new physics beyond the SM
without receiving any pollution from strong interactions. Numerical
results of such observables and comparisons with that obtained in
other theoretical methods, in particular QCDSR and PQCD approach,
are also included in this section. Section VI is devoted to our
conclusions.

\section{Effective Hamiltonian and parameterizations of matrix element}

\subsection{Effective Hamiltonian for $b \to s$ transition}

Integrating out the particles including top quark, $W^{\pm}$ and $Z$
bosons above scale $\mu=O(m_b)$ , we  arrive at the effective
Hamiltonian responsible for the $b \to s l^{+}l^{-}$ transition
\cite{H}
\begin{eqnarray}
H_{eff}(b\to s l^+l^-) &=& -\frac{G_{F}}{2\sqrt{2}}V_{tb}V_{ts}^{*}
\bigg[ {\sum\limits_{i=1}^{6}} C_{i}({\mu}) Q_{i}({\mu})+C_{7
\gamma}(\mu) Q_{7\gamma}(\mu)+C_{8G}(\mu) Q_{8G}(\mu)\nonumber\\
&&+C_{9}(\mu) Q_{9 }(\mu)+C_{10}(\mu) Q_{10}(\mu) \bigg],
\label{effective haniltonian 1}
\end{eqnarray}
where we have neglected the terms proportional to $V_{ub}V_{us}^{*}$
on account of $|V_{ub}V_{us}^{*}/V_{tb}V_{ts}^{*}|<0.02$ and the
complete list of the operators can be found in \cite{H}. In the SM,
Wilson coefficients $C_i$ at scale $\mu=m_b$ calculated in the naive
dimensional regularization (NDR) scheme  are collected in Table
\ref{wilson coefficient}.

\begin{table}[tb]
 \caption{Numerical values of Wilson coefficients in the NDR scheme
 at scale $\mu=m_b$ \cite{b to s in theory 21,Huang, c.q. geng 4}.} \label{wilson coefficient}
\begin{center}
\begin{tabular}{c|c|c|c|c|c|c|c|c|c}
  \hline
  \hline
 $C_1$& $C_2$ & $C_3$ & $C_4$ & $C_5$ & $C_6$ & $C_{7 \gamma}$ & $C_{8 G}$ & $C_{9}$ & $C_{10}$ \\
 \hline
 $-0.218$ & 1.107 & 0.011 & $-0.026$ & 0.007 & $-0.031$ & $-0.305$ & $-0.15$ & 4.344 & $-4.669$ \\
  \hline
  \hline
\end{tabular}
\end{center}
\end{table}

In terms of the Hamiltonian in Eq. (\ref{effective haniltonian 1}),
we can derive the free quark decay amplitude for $b \to s
l^{+}l^{-}$ process as
\begin{eqnarray}
M(b\to s l^+l^-) &=& \frac{G_{F}}{2\sqrt{2}}V_{tb}V_{ts}^{*}
{\alpha_{em}\over \pi}\bigg \{-{2i \over q^2}C_7^{eff}(\mu) \bar{s}
\sigma_{\mu\nu}q^{\nu}(m_b R+m_sL) b  \bar{l}\gamma^{\mu}l \nonumber
\\&&+ C_{9}^{eff}(\mu)\bar{s}\gamma_{\mu}Lb
\bar{l}\gamma^{\mu}l+C_{10}
\bar{s}\gamma_{\mu}Lb\bar{l}\gamma^{\mu}\gamma_5l \bigg\}. \label{b
to s l l}
\end{eqnarray}
We should emphasize that the Wilson coefficient $C_{10}$ does not
renormalize under QCD corrections and hence is independent on the
energy scale $\mu \simeq O(m_b)$, since the operator $O_{10}$ can
not be induced by the insertion of four-quark operators due to the
absence of $Z$ boson in the effective theory. Moreover, the above
quark decay amplitude can also receive additional contributions from
the matrix element of four-quark operators, $ \sum_{i=1}^{6}\langle
l^{+}l^{-}s| O_{i}| b \rangle$, which are usually absorbed into the
effective Wilson coefficient $C_9^{eff}(\mu)$. To be more specific,
we can decompose $C_9^{eff}(\mu)$ into the following three parts
\cite{b to s in theory 3,b to s in theory 4,b to s in theory 5,b to
s in theory 6,b to s in theory 7,b to s in theory 8,b to s in theory
9}
\begin{eqnarray}
C_9^{eff}(\mu) = C_9(\mu)+Y_{SD}(z,s')+Y_{LD}(z,s'),
\end{eqnarray}
where the parameters $z$ and $s'$ are defined as $z=m_c/m_b, \,\,\,
s'=q^2/m_b^2$. $Y_{SD}(z,s')$ describes the short-distance
contributions from four-quark operators far away from the $c\bar{c}$
resonance regions, which can be calculated reliably in perturbative
theory. The long-distance contributions $Y_{LD}(z,s')$ from
four-quark operators near the $c\bar{c}$ resonance cannot be
calculated from first principles of QCD and are usually
parameterized in the form of a phenomenological Breit-Wigner formula
making use of the vacuum saturation approximation and quark-hadron
duality. The manifest expressions for $Y_{SD}(z,s')$ and
$Y_{LD}(z,s')$ can be written as \cite{c.q. geng 1, c.q. geng 2,c.q.
geng 3,c.q. geng 4}
\begin{eqnarray}
Y_{SD}(z,s')&=&h(z,s')(3C_1(\mu)+C_2(\mu)+3C_3(\mu)+C_4(\mu)+3C_5(\mu)+C_6(\mu))\nonumber\\
&&-\frac{1}{2}h(1,s')(4C_3(\mu)+4C_4(\mu)+3C_5(\mu)
+C_6(\mu))\nonumber\\
&&-\frac{1}{2}h(0,s')(C_3(\mu)+3C_4(\mu))+{2 \over
9}(3C_3(\mu)+C_4(\mu)+3C_5(\mu) +C_6(\mu)),
\end{eqnarray}
\begin{eqnarray}
Y_{LD}(z,s')&=&\frac{3}{\alpha_{em}^{2}}(3C_1(\mu)+C_2(\mu)+3C_3(\mu)+C_4(\mu)+3C_5(\mu)+C_6(\mu))
\nonumber \\ && \sum_{j=\psi,\psi'}\omega_j(q^2)k_{j}
\frac{\pi\Gamma(j\rightarrow
l^{+}l^{-})M_{j}}{q^2-M_j^2+iM_{j}\Gamma_j^{tot}},\label{LD}
\end{eqnarray}
with
\begin{eqnarray}
h(z,s') &=& -{8\over 9}{\rm {ln}}z+{8\over 27}+{4\over 9}x-{2\over
9}(2+x)|1-x|^{1/2}
 \left\{
\begin{array}{l}
\ln \left| \frac{\sqrt{1-x}+1}{\sqrt{1-x}-1}\right| -i\pi \quad {\rm {for}}%
{ {\ }x\equiv 4z^{2}/s^{\prime }<1} \\
2\arctan \frac{1}{\sqrt{x-1}}\qquad {\rm {for}}{ {\ }x\equiv
4z^{2}/s^{\prime }>1}
\end{array}
\right., \nonumber\\
h(0,s^{\prime}) &=& {8 \over 27}-{8 \over 9} {\rm ln}{m_b \over \mu}
-{4 \over 9} {\rm {ln}}s^{\prime} +{4 \over 9}i \pi \,\, .
\end{eqnarray}
Here $M_j(\Gamma_j^{tot})$ are the masses (widths) of the
intermediate resonant states and $\Gamma(j\rightarrow l^{+}l^{-})$
denote the partial decay width for the transition of vector
charmonium state to massless lepton pair, which can be expressed in
terms of the decay constant of charmonium through the relation
\cite{b to s 1}
\begin{eqnarray}
\Gamma(j\rightarrow l^{+}l^{-})=\pi \alpha_{em}^2 {16  \over
27}{f_{j}^2 \over M_j}.
\end{eqnarray}
The phenomenological parameter $k_j$ in Eq.(\ref{LD}) is introduced
to account for inadequacies of the factorization approximation,
which can be determined from
\begin{eqnarray}
BR(\Lambda_b \to \Lambda J/\psi \to \Lambda l^{+}
l^{-})=BR(\Lambda_b \to \Lambda J/\psi) \cdot BR(J/\psi \to l^{+}
l^{-}).
\end{eqnarray}
The function $\omega_j(q^2)$  introduced in Eq.(\ref{LD}) is to
compensate the naive treatment of long distance contributions due to
the charm quark loop in the spirit of quark-hadron duality, which
can overestimate the genuine effect of the charm quark at small
$q^2$ remarkably \footnote {For a more detailed discussion on
long-distance and short-distance contributions from the charm loop,
one can refer to references \cite{b to s in theory 21,b to s 1, b to
s 2, b to s 3,charm loop 1, charm loop 2,charm loop 3}.}. The
quantity $\omega_j(q^2)$ can be normalized to
$\omega_j(M^2_{\psi_j}) = 1$, but its exact form is unknown at
present. It has been shown that there is a suppression for
$\omega_j(q^2)$ when going down to $q^2 = 0$ with $\omega(0) < 0.13$
(and could be even smaller) \cite{x.g. he LD}, hence we can neglect
the resonant contributions for $b \to s \gamma$ safely. Since the
dominant contribution of the resonances is in the vicinity of the
intermediate $\psi_i$ masses, we will simply use $\omega_j(q^2)=1$
in our numerical calculations. Moreover, the non-factorizable
effects \cite{b to s 1, b to s 2, b to s 3,NF charm loop} from the
charm loop can bring about further corrections to the radiative $b
\to s \gamma$ transition, which can be absorbed into the effective
Wilson coefficient $C_7^{eff}$ as usual. Specifically, the Wilson
coefficient $C^{eff}_{7}$ is given by \cite{c.q. geng 4}
\begin{eqnarray}
C_7^{eff}(\mu) = C_7(\mu)+C'_{b\to s\gamma}(\mu),
\end{eqnarray}
with
\begin{eqnarray}
C'_{b\rightarrow s \gamma}(\mu) &=& i\alpha_s \bigg[{2\over
9}\eta^{14/23}(G_1(x_t)-0.1687)-0.03C_2(\mu) \bigg], \\
G_1(x)  &=& {x(x^2-5x-2)\over 8(x-1)^3}+{3x^2 {\rm{ln}}^2x\over
4(x-1)^4},
\end{eqnarray}
where $\eta=\alpha_s(m_W)/\alpha_s(\mu)$, $x_t=m_t^2/m_W^2$,
$C'_{b\rightarrow s\gamma}$ is the absorptive part for the $b \to s
c \bar{c} \to s \gamma$ rescattering and we have dropped out the
tiny contributions proportion to CKM sector $V_{ub}V_{us}^{\ast}$.

Similarly, the free quark decay amplitude for $b\to s\gamma$ can be
written as \cite{H}:
\begin{eqnarray}
M(b\to s\gamma) &=& \frac{G_{F}}{2\sqrt{2}}V_{tb}V_{ts}^{*} {e\over
4\pi^2} C^{eff}_7(\mu) \bar{s} \sigma_{\mu\nu}q^{\nu}(m_b R+m_sL) b
F^{\mu\nu}. \label{b to s gamma}
\end{eqnarray}
We stress again that one should add a term
\begin{eqnarray}
\left[3C_1(\mu)+C_2(\mu)+3C_3(\mu)+C_4(\mu)+3C_5(\mu)+C_6(\mu)\right]\frac{3}{\alpha_{em}^{2}}\sum_{j=\psi,\psi'}
\omega_j(0)k_j\frac{\pi\Gamma(j\rightarrow
l^{+}l^{-})M_{j}}{q^2-M_j^2+iM_{j}\Gamma_j^{tot}}
\end{eqnarray}
to the effective Wilson coefficient $C^{eff}_7$, if the
long-distance contributions from the charm quark loop  in the
resonance regions are included. As mentioned above, this type of
effects are suppressed heavily both by  the Breit-Wigner factor
$\sim \Gamma_i/M_i$ and phenomenological parameter $\omega(0)$, and
hence are neglected in the later discussions.

\subsection{Parameterizations of hadronic matrix element}
\label{Parameterizations of hadronic matrix element}

With the free quark decay amplitude available, we can proceed to
calculate the decay amplitudes for $\Lambda_b\to \Lambda \gamma$ and
$\Lambda_b \to \Lambda l^+l^-$ at hadron level, which can be
obtained by sandwiching the free quark amplitudes between the
initial and final baryon states. Consequently, the following four
hadronic matrix elements
\begin{eqnarray}
\langle \Lambda(P)|\bar{s}\gamma_{\mu} b|\Lambda_{b}(P+q)\rangle
&,& \,\,\,  \langle \Lambda(P)|\bar{s}\gamma_{\mu}\gamma_5 b|\Lambda_{b}(P+q)\rangle , \nonumber \\
\langle \Lambda(P)|\bar{s}\sigma_{\mu \nu} b|\Lambda_{b}(P+q)\rangle
&,& \,\,\, \langle \Lambda(P)|\bar{s}\sigma_{\mu \nu} \gamma_5
b|\Lambda_{b}(P+q)\rangle,
\end{eqnarray}
need to be computed as can be observed from Eqs. (\ref{b to s l l})
and (\ref{b to s gamma}). Generally, the above four matrix elements
can be parameterized in terms of a series of form factors as
\cite{c.q. geng 4, Aliev 1,Aliev 2,Aliev 3,Aliev 4}
\begin{eqnarray}
\langle \Lambda(P)|\bar{s}\gamma_{\mu} b|\Lambda_{b}(P+q)\rangle
&=&\overline{\Lambda}(P)(g_1 \gamma_{\mu}+g_2 i \sigma_{\mu \nu}
q^{\nu}+g_3 q_{\mu})\Lambda_b(P+q), \,\, \label{vector
matrix element}\\
\langle \Lambda(P)|\bar{s}\gamma_{\mu}\gamma_5
b|\Lambda_{b}(P+q)\rangle &=&\overline{\Lambda}(P)(G_1
\gamma_{\mu}+G_2 i\sigma_{\mu \nu} q^{\nu}+G_3
q_{\mu})\gamma_{5}\Lambda_b(P+q), \,\, \label{axial-vector
matrix element}\\
\langle\Lambda(P)|\bar{s}\sigma_{\mu \nu} b|\Lambda_b(P+q)\rangle
&=&\overline{\Lambda}(P)[h_1 \sigma_{\mu \nu} -i h_2
(\gamma_{\mu}q_{\nu}-\gamma_{\nu}q_{\mu} )\nonumber \\
&&-i h_3 (\gamma_{\mu}P_{\nu}-\gamma_{\nu}P_{\mu} ) -i h_4
(P_{\mu}q_{\nu}-P_{\nu}q_{\mu})]\Lambda_b(P+q), \,\,
\label{tensor matrix element 1}\\
\langle\Lambda(P)|\bar{s}\sigma_{\mu \nu}
\gamma_{5}b|\Lambda_b(P+q)\rangle &=&\overline{\Lambda}(P) [H_1
\sigma_{\mu \nu} -i H_2
(\gamma_{\mu}q_{\nu}-\gamma_{\nu}q_{\mu} )\nonumber \\
&&-i H_3 (\gamma_{\mu}P_{\nu}-\gamma_{\nu}P_{\mu} ) -i H_4
(P_{\mu}q_{\nu}-P_{\nu}q_{\mu})] \gamma_5 \Lambda_b(P+q),
\label{tensor matrix element 2}
\end{eqnarray}
where all the form factors $g_i$, $G_i$, $h_i$ and $H_i$ are
functions of the square of momentum transfer $q^2$. Rewriting the
Eqs. (\ref{tensor matrix element 1}-\ref{tensor matrix element 2}),
we have the baryon matrix elements for the dipole operators as
following
\begin{eqnarray}
\langle\Lambda(P)|\bar{s}i \sigma_{\mu \nu} q^{\nu}
b|\Lambda_b(P+q)\rangle &=&\overline{\Lambda}(P)(f_1
\gamma_{\mu}+f_2 i \sigma_{\mu \nu} q^{\nu}+f_3
q_{\mu})\Lambda_b(P+q),
\,\,\label{magnetic matrix element 1}\\
\langle\Lambda(P)|\bar{s}i \sigma_{\mu \nu}\gamma_{5} q^{\nu}
b|\Lambda_b(P+q)\rangle &=&\overline{\Lambda}(P)(F_1
\gamma_{\mu}+F_2 i \sigma_{\mu \nu} q^{\nu}+F_3
q_{\mu})\gamma_{5}\Lambda_b(P+q), \label{magnetic matrix element 2}
\end{eqnarray}
with
\begin{eqnarray}
f_1&=& {2h_2-h_3+h_4(m_{\Lambda_b}+m_{\Lambda}) \over 2} q^2 , \\
f_2&=&  { 2 h_1 +h_3(m_{\Lambda}-m_{\Lambda_b}) +h_4 q^2 \over 2} , \\
f_3&=& { m_{\Lambda}-m_{\Lambda_b} \over q^2}f_1  , \\
F_1&=&  {2H_2-H_3+H_4(m_{\Lambda_b}-m_{\Lambda}) \over 2} q^2 , \\
F_2&=&  { 2 H_1 +H_3(m_{\Lambda}+m_{\Lambda_b}) +H_4 q^2 \over 2}  , \\
F_3&=&  { m_{\Lambda}+m_{\Lambda_b} \over q^2}F_1  .
\end{eqnarray}

It should be emphasized that the form factors $f_3$ and $F_3$ do not
contribute to the decay amplitude of $\Lambda_b \to \Lambda + l^{+}
l^{-}$ due to the conservation of vector current, namely  $q^{\mu}
\bar{l} \gamma_{\mu} l = 0$.  Concentrating on the radiative decay
of $\Lambda_b\to \Lambda \gamma$, we then observe that the matrix
element of magnetic penguin operators can be simplified as
\begin{eqnarray}
\label{parameterization of tensor current 1}
\langle\Lambda(P)|\bar{s}i \sigma_{\mu \nu} q^{\nu}
b|\Lambda_b(P+q)\rangle &=& f_2(0) \overline{\Lambda}(P) i
\sigma_{\mu \nu} q^{\nu}\Lambda_b(P+q),
\,\,\\
\langle\Lambda(P)|\bar{s}i \sigma_{\mu \nu}\gamma_{5} q^{\nu}
b|\Lambda_b(P+q)\rangle &=&F_2(0) \overline{\Lambda}(P) i
\sigma_{\mu \nu}\gamma_5  q^{\nu}\Lambda_b(P+q).
\label{parameterization of tensor current 2}
\end{eqnarray}

For the completeness, we also present the parameterizations of
matrix elements involving the scalar $\bar{s} b$ and pseudo-scalar
$\bar{s} \gamma_5 b$ currents, which can be obtained from the Eqs.
(\ref{vector matrix element}) and (\ref{axial-vector matrix
element}) by contracting both sides to the four-momentum $q^{\mu}$
\begin{eqnarray}
\langle\Lambda(P)|\bar{s} b|\Lambda_b(P+q)\rangle &=& {1 \over
m_b+m_s} \overline{\Lambda}(P) [g_1(m_{\Lambda_b}-m_{\Lambda})+g_3
q^2]\Lambda_b(P+q), \,\, \label{scalar
matrix element}\\
\langle\Lambda(P)|\bar{s}\gamma_{5} b|\Lambda_b(P+q)\rangle &=& {1
\over m_b-m_s} \overline{\Lambda}(P)
[G_1(m_{\Lambda_b}+m_{\Lambda})-G_3 q^2]\gamma_5\Lambda_b(P+q).
\label{pseudo-scalar matrix element}
\end{eqnarray}
The   two independent form factors $\xi_1$ and  $\xi_2$ in HQET  are
defined as
\begin{eqnarray}
\langle\Lambda(P)|\bar{b}\Gamma s|\Lambda_{b}(P+q)\rangle
&=&\overline{\Lambda}(P)[\xi_1(q^2)+\not \! v \xi_2(q^2)] \Gamma
\Lambda_b(P+q), \label{form factors in HQET}
\end{eqnarray}
with $\Gamma$ being an arbitrary Lorentz structure and $v_{\mu}$
being the four-velocity of $\Lambda_b$ baryon. Comparing Eqs.
(\ref{vector matrix element}-\ref{axial-vector matrix element}),
(\ref{magnetic matrix element 1}-\ref{magnetic matrix element 2})
and the Eq. (\ref{form factors in HQET}), one can easily find
  \cite{c.q. geng 4, Aliev 1,Aliev 2,Aliev 3,Aliev 4}
\begin{eqnarray}
f_1&=& F_1 = {q^2 \over m_{\Lambda_b}} \xi_2, \\
f_2&=&F_2=g_1=G_1= \xi_1 + { m_{\Lambda} \over  m_{\Lambda_b}}\xi_2,
\label{relatins of from factors in HQET 1} \\
f_3&=&  {m_{\Lambda}-m_{\Lambda_b }\over m_{\Lambda_b}} \xi_2 , \\
F_3&=&  {m_{\Lambda}+m_{\Lambda_b }\over m_{\Lambda_b}} \xi_2 , \\
g_2&=&G_2=g_3=G_3={ \xi_2 \over  m_{\Lambda_b}}. \label{relatins of
from factors in HQET 2}
\end{eqnarray}
It is known that  Eq. (\ref{form factors in HQET}) is successful at
zero recoil region (with  large  $q^2$) in the heavy quark limit. As
for the large recoil region, the large energy effective theory
implies that the form factors are independent of both energy of
light hadron ($E$) and the heavy quark mass ($m_b$) with the
assumption of Feynman mechanism, which indicates that Eq. (\ref{form
factors in HQET}) is still well defined owing to the tiny effects
from ${1/ m_b}$, ${ 1 / E}$ and $\alpha_s$ corrections
\cite{Hiller:2001zj}.

\section{Light-cone sum rules for the transition form factors}


The most general decomposition of matrix element for three-quark
operator with light-like separations $x^2 \to 0$ between vacuum and
$\Lambda$ baryon state at tree level can be written as \cite{DAs of
nucleon 1}
\begin{eqnarray}
&&4\langle 0 |\epsilon^{ijk} u_{\alpha}^{i}(a_1 x) d_{\beta}^{j}(a_2
x) s_{\gamma}^{k}(a_3 x)|\Lambda(P)\rangle \nonumber \\&&= {\cal
S}_1 m_{\Lambda} C_{\alpha \beta} \left(\gamma_5
\Lambda\right)_\gamma + {\cal S}_2 m_{\Lambda}^2 C_{\alpha \beta}
\left(\!\not\!{x} \gamma_5 \Lambda\right)_\gamma + {\cal P}_1
m_{\Lambda} \left(\gamma_5 C\right)_{\alpha \beta} \Lambda_\gamma +
{\cal P}_2 m_{\Lambda}^2 \left(\gamma_5 C \right)_{\alpha \beta}
\left(\!\not\!{x} \Lambda\right)_\gamma
\nonumber \\
&& + {\cal V}_1  \left(\!\not\!{P}C \right)_{\alpha \beta}
\left(\gamma_5 \Lambda\right)_\gamma + {\cal V}_2 m_{\Lambda}
\left(\!\not\!{P} C \right)_{\alpha \beta} \left(\!\not\!{x}
\gamma_5 \Lambda\right)_\gamma  + {\cal V}_3 m_{\Lambda}
\left(\gamma_\mu C \right)_{\alpha \beta}\left(\gamma^{\mu} \gamma_5
\Lambda\right)_\gamma
\nonumber \\
&& + {\cal V}_4 m_{\Lambda}^2 \left(\!\not\!{x}C \right)_{\alpha
\beta} \left(\gamma_5 \Lambda\right)_\gamma + {\cal V}_5
m_{\Lambda}^2 \left(\gamma_\mu C \right)_{\alpha \beta} \left(i
\sigma^{\mu\nu} x_\nu \gamma_5 \Lambda\right)_\gamma + {\cal V}_6
m_{\Lambda}^3 \left(\!\not\!{x} C \right)_{\alpha \beta}
\left(\!\not\!{x} \gamma_5 \Lambda\right)_\gamma
\nonumber \\
&& + {\cal A}_1  \left(\!\not\!{P}\gamma_5 C \right)_{\alpha \beta}
\Lambda_\gamma + {\cal A}_2 m_{\Lambda} \left(\!\not\!{P}\gamma_5 C
\right)_{\alpha \beta} \left(\!\not\!{x} \Lambda\right)_\gamma  +
{\cal A}_3 m_{\Lambda} \left(\gamma_\mu \gamma_5 C \right)_{\alpha
\beta}\left( \gamma^{\mu}\Lambda\right)_\gamma
\nonumber \\
&& + {\cal A}_4 m_{\Lambda}^2 \left(\!\not\!{x} \gamma_5 C
\right)_{\alpha \beta} \Lambda_\gamma + {\cal A}_5 m_{\Lambda}^2
\left(\gamma_\mu \gamma_5 C \right)_{\alpha \beta} \left(i
\sigma^{\mu\nu} x_\nu \Lambda\right)_\gamma + {\cal A}_6
m_{\Lambda}^3 \left(\!\not\!{x} \gamma_5 C \right)_{\alpha \beta}
\left(\!\not\!{x} \Lambda\right)_\gamma
\nonumber \\
&& + {\cal T}_1 \left(P^\nu i \sigma_{\mu\nu} C\right)_{\alpha
\beta} \left(\gamma^\mu\gamma_5 \Lambda\right)_\gamma + {\cal T}_2
m_{\Lambda} \left(x^\mu P^\nu i \sigma_{\mu\nu} C\right)_{\alpha
\beta} \left(\gamma_5 \Lambda\right)_\gamma +{\cal T}_3 m_{\Lambda}
\left(\sigma_{\mu\nu} C\right)_{\alpha \beta}
\left(\sigma^{\mu\nu}\gamma_5 \Lambda\right)_\gamma
\nonumber \\
&& + {\cal T}_4 m_{\Lambda} \left(P^\nu \sigma_{\mu\nu}
C\right)_{\alpha \beta} \left(\sigma^{\mu\rho} x_\rho \gamma_5
\Lambda\right)_\gamma + {\cal T}_5 m_{\Lambda}^2 \left(x^\nu i
\sigma_{\mu\nu} C\right)_{\alpha \beta} \left(\gamma^\mu\gamma_5
\Lambda\right)_\gamma
\nonumber \\
&& + {\cal T}_6 m_{\Lambda}^2 \left(x^\mu P^\nu i \sigma_{\mu\nu}
C\right)_{\alpha \beta} \left(\!\not\!{x} \gamma_5
\Lambda\right)_\gamma + {\cal T}_{7} m_{\Lambda}^2
\left(\sigma_{\mu\nu} C\right)_{\alpha \beta} \left(\sigma^{\mu\nu}
\!\not\!{x} \gamma_5 \Lambda\right)_\gamma
\nonumber \\
&&+ {\cal T}_{8} m_{\Lambda}^3 \left(x^\nu \sigma_{\mu\nu}
C\right)_{\alpha \beta} \left(\sigma^{\mu\rho} x_\rho \gamma_5
\Lambda\right)_\gamma \,, \label{DAs 1}
\end{eqnarray}
in view of the Lorentz covariance, spin and parity of $\Lambda$
baryon. Here, each of the 24 invariant functions $\mathcal{F}_i,
(\mathcal{F}=\mathcal{S},\mathcal{P},\mathcal{V},\mathcal{A},\mathcal{T})$
is a function of  the scalar product $P \cdot x$ and the parameters
$a_i$ denote coordinates of valence quarks.

The ``calligraphic" functions $\mathcal{F}_i$ defined above do not
have a definite twist. Decomposing the quark field operators into
``plus" and ``minus" components  \cite{DAs of nucleon 1} and
utilizing the equation of motion, the invariant function
$\mathcal{F}_i$ can be expressed by the distribution amplitudes
$A_i$ \cite{m.q. huang}:
\begin{eqnarray}
{\cal{A}}_1 = A_1\,, &~~&   2 p\cdot x  {\cal{A}}_2 = - A_1 + A_2 -  A_3\,, \nonumber \\
2 {\cal{A}}_3 = A_3\,, &~~&  4 p\cdot x  {\cal{A}}_4 = - 2 A_1 - A_3 - A_4  + 2 A_5\,, \nonumber \\
4 p\cdot x  {\cal{A}}_5 = A_3 - A_4\,, &~~& (2 p\cdot x )^2
{\cal{A}}_6 =  A_1 - A_2 +  A_3 +  A_4 - A_5 + A_6\,,
\label{relations between two types of DAs}
\end{eqnarray}
where the twists of distribution amplitudes $A_i(i=1,...,6)$ can be
easily determined by power countering as  collected in
Table~\ref{twist}. Only the the relations associated with
axial-vector distribution amplitudes $A_i$ of $\Lambda$ baryon are
involved in the sum rules of form factors responsible for $\Lambda_b
\to \Lambda$ transition, since the spectator quarks $[ud]$ in the
$\Lambda_b$ baryon maintain their position at the origin up to the
accuracy of leading order in $\alpha_s$.

\begin{table}[tb]
\caption{Twist classification \cite{DAs of nucleon 1} of the
distribution amplitudes given in Eq.~(\ref{relations between two
types of DAs}). }
\begin{center}
\begin{tabular}{l l l l l}
\hline
\hline &  \hspace{2 cm} twist-3  & \hspace{2 cm} twist-4  & \hspace{2 cm} twist-5   &  \hspace{2 cm} twist-6  \\
 axial-vector & \hspace{2 cm} $A_1$    & \hspace{2 cm} $A_2\;,\;A_3$ & \hspace{2 cm} $A_4\;,\;A_5 $& \hspace{2 cm} $A_6$ \\
\hline \hline
\end{tabular}
\end{center}
\label{twist}
\end{table}


Generally, the distribution amplitudes depend on the renormalization
scale and can be expanded in contributions of light-ray operators
with increasing conformal spin, where the constraints on operator
mixing and equation of motion owing to conformal symmetry of the QCD
Lagrangian \cite{CFS} are also taken into account. To the leading
logarithmic accuracy, the different conformal partial waves do not
mix with each other, since the renormalization group (RG) are driven
by tree-level counter terms and the conformal symmetry is well
respected. Besides, it is known that QCDSR tends to overestimate the
matrix elements corresponding to the higher conformal spin operators
considerably from the experience of both calculating the pion
electromagnetic form factor \cite{form factor of pion 1,form factor
of pion 2} and nucleon form factor \cite{form factor of nucleon}.
 Therefore, following the Ref. \cite{m.q. huang}, we would  like
to adopt the distribution amplitudes of $\Lambda$ baryon to the
leading conformal spin accuracy in the present work.

The explicit forms of distribution amplitudes for $\Lambda$ baryon
to the accuracy of leading conformal spin can be expressed as
\cite{m.q. huang}
\begin{eqnarray}
A_1(x_1, x_2,x_3)&=&-120 x_1 x_2 x_3 \phi_3^0,  \nonumber \\
A_2(x_1, x_2,x_3)&=& -24 x_1 x_2 \phi_4^0,  \nonumber \\
A_3(x_1, x_2,x_3)&=& -12 x_3  (1-x_3) \psi_4^0,  \nonumber \\
A_4(x_1, x_2,x_3)&=& -3 (1-x_3) \phi_5^0,  \nonumber \\
A_5(x_1, x_2,x_3)&=&-6 x_3 \phi_5^0,  \nonumber \\
A_6(x_1, x_2,x_3)&=&-2 \phi_6^0, \label{asymptonic forms of A_i}
\end{eqnarray}
with
\begin{eqnarray}
\phi_3^0=\phi_6^0=-f_{\Lambda},\, \,\, \, \,
\phi_4^0=\phi_5^0=-{f_{\Lambda}+\lambda_1 \over 2},\, \,\,\,
\psi_4^0=\psi_5^0={f_{\Lambda}-\lambda_1 \over 2}\,.
\end{eqnarray}
The estimations of non-perturbative parameters $f_{\Lambda}$ and
$\lambda_1$ in the framework of QCDSR approach have been presented
in Ref. \cite{m.q. huang} and to make the paper self contained,
these are collected in the Appendix \ref{decay constants of lambda}.

With the LCDAs of $\Lambda$ baryon available, we are now in a
position to derive the sum rules of transition form factors which
are responsible for $\Lambda_b \to \Lambda \gamma$ and $\Lambda_b
\to \Lambda l^{+} l^{-}$ decays. The basic object in LCSR approach
is the correlation function   where one of the hadron is represented
by the interpolating current with proper quantum number, such as
 spin, isospin, (charge) parity  and so on; and the
other is described by its vector state manifestly. Information on
the hadronic transition form factor can be extracted by matching the
Green function calculated in two different representations, i.e.,
phenomenological and theoretical forms, with the help of  dispersion
relation under the assumption of quark-hadron duality.

\subsection{Light-cone sum rules for $\Lambda_b \to \Lambda +
\gamma$}

We consider the correlation function associating with $\Lambda_b \to
\Lambda + \gamma$ decay determined by the matrix element
\begin{eqnarray}
z^{\nu} T_{\nu}(P,q)=i z^{\nu} \int d^4 x e^{-i q \cdot x} \langle 0
|T \{ j_{\Lambda_b}(0) j_{\nu}(x)\} | \Lambda  (P) \rangle,
\label{correlation function tensor current}
\end{eqnarray}
between the vacuum and $\Lambda$ baryon state $| \Lambda  (P)
\rangle$.  Generally, choice of interpolating current for baryon, in
particular the light baryon, is not unique \cite{choice of currect
1,choice of currect 2,choice of currect 3,choice of currect 4,choice
of currect 5,choice of currect 6}. The interpolating  field
$j_{\Lambda_b}(0)$ denoting the $\Lambda_b$ baryon and the
transition current $j_{\nu}(x)$ describing the transition of
$\Lambda_ b \to \Lambda \gamma$ are given by
\begin{eqnarray}
j_{\Lambda_b}(0)&=&\epsilon^{ijk}[u^{i}(0)C\gamma_5 \not \!z d^j(0)]
\not \! z b^k(0) \,\,, \nonumber \\
j_{\nu}(x)&=& i  \bar{b}(x) \sigma_{\mu
\nu}(1-\gamma_5)q^{\mu}s(x)\,\,. \label{choice of current}
\end{eqnarray}
In addition, the vacuum-to-baryon matrix element for the
interpolating current can be parameterized as follows
\begin{eqnarray}
\langle 0 | j_{\Lambda_b}(0) | \Lambda_b(P^{\prime})  \rangle
=f_{\Lambda_b} (z \cdot P^{\prime}) \not \!z \Lambda_b(P^{\prime}).
\label{coupling of lambda_b}
\end{eqnarray}
Inserting the complete set of states between the currents in Eq.
(\ref{correlation function tensor current}) with the same quantum
numbers as $\Lambda_b$, we can arrive at the hadronic representation
of the correlator
\begin{eqnarray}
z^{\nu} T_{\nu}={\langle 0 | j_{\Lambda_b} | \Lambda_b(P^{\prime})
\rangle   \langle\Lambda_b(P^{\prime}) |z^{\nu} j_{\nu} |
\Lambda(P)\rangle \over m_{\Lambda_b}^2 -{P^{\prime}}^2} + \sum_h
{\langle 0 | j_{\Lambda_b} | h(P^{\prime}) \rangle \langle
h(P^{\prime}) |z^{\nu} j_{\nu} | \Lambda(P)\rangle \over m_{h}^2
-{P^{\prime}}^2}, \label{inserting complete states}
\end{eqnarray}
where $P^{\prime}=P+q$ and we have separated the  contributions from
the ground state and higher states corresponding to the $\Lambda_b$
baryon channel. Combining the Eq. (\ref{magnetic matrix element 1},
\ref{magnetic matrix element 2}), (\ref{coupling of lambda_b},
\ref{inserting complete states}) and summing over the polarization
of $\Lambda_b$ baryon, we can obtain the phenomenological
representations of correlation function as
\begin{eqnarray}
z^{\nu} T_{\nu}=2 f_{\Lambda_b} {(z \cdot P^{\prime})^2 \over
m_{\Lambda_b}^2-{P^{\prime}}^2} \bigg[-f_1 \not \! z +f_2 \not \! z
\not \! q - F_1 \not \! z \gamma_5  - F_2 \not \! z \not \! q
\gamma_5 \bigg] \Lambda(P)+..., \label{correlation function at
hadron level}
\end{eqnarray}
where the ellipses stand for the terms   proportional to the  higher
power of $1/P$ in the infinite momentum kinematics $P\sim \infty,
\,\, q \sim const, \,\, z \sim 1/P$ and the contributions from the
higher states of $\Lambda_b$ channel. The advantage of contracting
the correlation function with $z^{\nu}$ is that the Lorentz
structures can be simplified   owing to the null contributions from
$~ z_{\nu}$ on the light-cone.

\begin{figure}[tb]
\begin{center}
\begin{tabular}{ccc}
\includegraphics[scale=0.6]{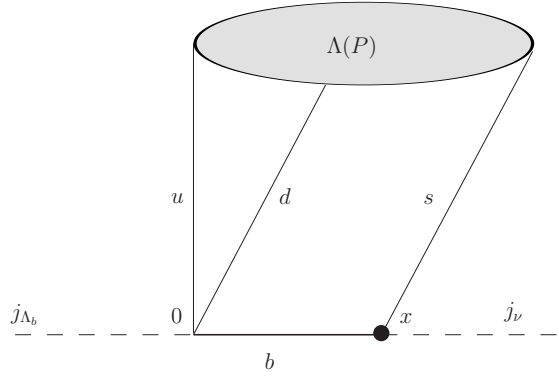}
\end{tabular}
\caption{The tree level contribution to the correlation function
(\ref{correlation function tensor current}), where the thick solid
line represents the heavy $b$ quark.} \label{transition}
\end{center}
\end{figure}

On the theoretical side,  the correlation function (\ref{correlation
function tensor current}) can also be computed in the perturbation
theory with the help of OPE technique at the deep Euclidean region:
${P^{\prime}}^2, \,\, q^2=-Q^2 \ll 0$. To the leading order of
$\alpha_s$, the correlation function can be calculated by
contracting the bottom quark field in Eq. (\ref{correlation function
tensor current}) and inserting the free $b$ quark propagator
\begin{eqnarray}
z^{\nu} T_{\nu}=-2 (C \gamma_5 \not \! z)_{\alpha \beta} [\not \! z
(1-\gamma_5)]_{\gamma} \int d^4 x \int {d^4 k \over (2 \pi)^4}e^{i
(k-q) \cdot x} {z \cdot k \over k^2-m_b^2}  \langle 0
|\epsilon^{ijk} u_{\alpha}^{i}(0) d_{\beta}^{j}(0)
s_{\gamma}^{k}(x)|\Lambda(P)\rangle, \label{correlation function at
quark level tensor current}
\end{eqnarray}
shown in Fig. \ref{transition}.

The full quark propagator also receives corrections from the
background field \cite{asymptonic forms 1,Khodjamirian:1998ji} and
can be written as
\begin{eqnarray}
\langle 0| T \{{b_i(x) \bar{b}_j(0)}\}| 0\rangle &=& \delta_{ij}\int
{d^4 k \over (2 \pi)^4} e^{-i kx}{i \over \not \! k -m_b} -i g \int
{d^4 k \over (2 \pi)^4} e^{-i kx} \int_0^1 dv  [{1 \over 2} {\not k
+m_b \over (m_b^2 -k^2)^2} G^{\mu \nu}_{ij}(v x)\sigma_{\mu \nu }\nonumber \\
&& +{1 \over m_b^2-k^2}v x_{\mu} G^{\mu \nu}(v x)\gamma_{\nu}],
\end{eqnarray}
where the first term is the free-quark propagator and $G^{\mu
\nu}_{i j}=G_{\mu \nu}^{a} T^a_{ij}$ with ${\mbox{Tr}}[T^a T^b]={1
\over 2}\delta^{ab}$. Inserting the second term  proportional to the
gluon field strength into the correlation function can result in the
distribution amplitudes corresponding to the higher Fock states of
$\Lambda$ baryon. It is expected that such corrections associating
with the LCDAs of higher Fock states do not play any significant
role in the sum rules for transition form factors \cite{higher Fock
state}, and therefore can be neglected safely in the presented work.

Substituting Eq.~(\ref{DAs 1}) into Eq.~(\ref{correlation function
at quark level tensor current}) and performing the integral in the
coordinate space, we can achieve the correlation function in the
momentum representation at the quark level as
\begin{eqnarray}
z^{\nu} T_{\nu}&&=-2(z \cdot P)^2 \not \! z \not \! q
(1-\gamma_5)\Lambda(P) \bigg \{ \int {\cal{D}} x {x_3 A_1
(x_i) \over (x_3P+q)^2-m_b^2} \nonumber \\
&& \hspace{1.0 cm}+M^2 \int_0^1 dx_3 \,
x_3^2{-2\tilde{A}_1(x_3)+\tilde{A}_2(x_3)-\tilde{A}_3(x_3)
-\tilde{A}_4(x_3)+\tilde{A}_5(x_3)\over [(x_3P+q)^2-m_b^2]^2}\nonumber \\
&& \hspace{1.0 cm}+2 M^4 \int_0^1 dx_3 \, x_3^3{\tilde{\tilde
A}_1(x_3)-\tilde{\tilde A}_2(x_3)+\tilde{\tilde
A}_3(x_3)+\tilde{\tilde A}_4(x_3)-\tilde{\tilde
A}_5(x_3)+\tilde{\tilde A}_6(x_3)\over [(x_3P+q)^2-m_b^2]^3}\bigg\}\nonumber \\
&& +2 q^2 (z \cdot P)^2  \not \! z (1+\gamma_5)\Lambda(P)\bigg\{ M
\int_0^1 dx_3 \, x_3
{-\tilde{A}_1(x_3)+\tilde{A}_2(x_3)-\tilde{A}_3(x_3) \over
[(x_3P+q)^2-m_b^2]^2 } \nonumber \\
&&  +2 M^3 \int_0^1 dx_3 \, x_3^2{\tilde{\tilde
A}_1(x_3)-\tilde{\tilde A}_2(x_3)+\tilde{\tilde
A}_3(x_3)+\tilde{\tilde A}_4(x_3)-\tilde{\tilde
A}_5(x_3)+\tilde{\tilde A}_6(x_3)\over [(x_3P+q)^2-m_b^2]^3}
\bigg\}+...., \label{correlation function theoretically}
\end{eqnarray}
where subleading terms in the infinite momentum frame kinematics
denoted by the ellipses are not included. The distribution
amplitudes with tildes  are defined as
\begin{eqnarray}
\tilde{F}(x_3)&=& \int_1^{x_3} dx_3^{\prime} \int_0^{1-x_3^{\prime}}
dx_1 F(x_1,1-x_1-x_3^{\prime},x_3^{\prime})\,\,, \nonumber \\
\tilde{\tilde{F}}(x_3)&=&\int_1^{x_3} dx_3^{\prime}
\int_1^{x_3^{\prime}} dx_3^{\prime \prime} \int_0^{1-x_3^{\prime}}
dx_1 F(x_1,1-x_1-x_3^{\prime \prime},x_3^{\prime \prime}),
\end{eqnarray}
originating from the partial integral in the variable $x_3$ in order
to eliminate the factor $1/P \cdot x$ due to the insertion of
distribution amplitudes in Eq. (\ref{relations between two types of
DAs}).

For the convenience of matching the correlation in QCD
representation and hadronic level, the Eq. (\ref{correlation
function theoretically}) is usually written in a form of dispersion
integral as
\begin{eqnarray}
z^{\nu}T_{\nu}=(z \cdot P)^2 \int_{m_b^2}^{\infty}ds {\rho_{V}(s,
Q^2) \not \! z (1-\gamma_5) +\rho_{T}(s, Q^2) \not \! z \not \! q
(1-\gamma_5) \over s-{P^{\prime}}^2}  \Lambda(P)+...,
\label{dispersion integral}
\end{eqnarray}
where $Q^2$ is defined as $Q^2=-q^2$. With the assumption of
quark-hadron duality, the higher states in the $\Lambda_b$ channel
can  be  given by the same dispersion integral  only with the lower
bound replaced by the effective threshold parameter $s_0$.
 Besides, the Borel transformation is commonly
introduced in the standard procedure of sum rules approach for the
sake of compensating the deficiency due to the approximation of
quark-hadron duality.
Finally, we can  obtain the sum rules for the transition form
factors of   the $\Lambda_b \to \Lambda \gamma$ decay as
\begin{eqnarray}
f_{\Lambda_b} f_1(q^2) e^{-m_{\Lambda_b}^2 / M_B^2}&=& - q^2
M\bigg\{ \int_{x_0}^1 {dx_3 \over x_3}e^{-s/M_B^2} \bigg[ {1 \over
M_B^2}
\bigg(-\tilde{A}_1(x_3)+\tilde{A}_2(x_3)-\tilde{A}_3(x_3)\bigg)\nonumber \\
&& \hspace{1.0 cm}-{M^2 \over M_B^4}\bigg(\tilde{\tilde
A}_1(x_3)-\tilde{\tilde A}_2(x_3)+\tilde{\tilde
A}_3(x_3)+\tilde{\tilde A}_4(x_3)-\tilde{\tilde
A}_5(x_3)+\tilde{\tilde A}_6(x_3)\bigg)\bigg] \nonumber \\
&& +{x_0 e^{-s_0/M_B^2}\over m_b^2-q^2+x_0^2M^2} \bigg[ \bigg(
-\tilde{A}_1(x_3)+\tilde{A}_2(x_3)-\tilde{A}_3(x_3)\bigg)\nonumber \\
&&\hspace{1.0 cm}-{M^2 \over M_B^2} \bigg( \tilde{\tilde
A}_1(x_0)-\tilde{\tilde A}_2(x_0)+\tilde{\tilde
A}_3(x_0)+\tilde{\tilde A}_4(x_0)-\tilde{\tilde
A}_5(x_0)+\tilde{\tilde A}_6(x_0)\bigg) \label{sum relus for f1}\\
&& +M^2 x_0 {d \over dx_0} \Bigg({x_0  \bigg( \tilde{\tilde
A}_1(x_0)-\tilde{\tilde A}_2(x_0)+\tilde{\tilde
A}_3(x_0)+\tilde{\tilde A}_4(x_0)-\tilde{\tilde
A}_5(x_0)+\tilde{\tilde A}_6(x_0)\bigg)\over m_0^2-q^2+x_0^2M^2}
\Bigg)\bigg]\bigg\}\,\,, \nonumber
\end{eqnarray}
and
\begin{eqnarray}
f_{\Lambda_b} f_2(q^2) e^{-m_{\Lambda_b}^2 / M_B^2}&=& \int_{x_0}^1
dx_3 e^{-s/M_B^2} \bigg[ \bigg( \int_0^{1-x_3}dx_1
A_1(x_1,1-x_1-x_3,x_3)\bigg)\nonumber \\
&& \hspace{1.0 cm}-{M^2 \over
M_B^2}\bigg(-2\tilde{A}_1(x_3)+\tilde{A}_2(x_3)-\tilde{A}_3(x_3)
-\tilde{A}_4(x_3)+\tilde{A}_5(x_3)\bigg)\nonumber \\
&& \hspace{1.0 cm}+{M^4 \over M_B^4}\bigg( \tilde{\tilde
A}_1(x_3)-\tilde{\tilde A}_2(x_3)+\tilde{\tilde
A}_3(x_3)+\tilde{\tilde A}_4(x_3)-\tilde{\tilde
A}_5(x_3)+\tilde{\tilde A}_6(x_3)\bigg)\bigg] \nonumber \\
&& -{M^2 x_0^2 e^{-s_0/M_B^2}\over m_b^2-q^2+x_0^2M^2} \bigg[ \bigg(
-2\tilde{A}_1(x_0)+\tilde{A}_2(x_0)-\tilde{A}_3(x_0)
-\tilde{A}_4(x_0)+\tilde{A}_5(x_0)\bigg)\nonumber \\
&&\hspace{1.0 cm}-{M^2 \over M_B^2} \bigg( \tilde{\tilde
A}_1(x_0)-\tilde{\tilde A}_2(x_0)+\tilde{\tilde
A}_3(x_0)+\tilde{\tilde A}_4(x_0)-\tilde{\tilde
A}_5(x_0)+\tilde{\tilde A}_6(x_0)\bigg)\label{sum relus for f2} \\
&& +M^2 {d \over dx_0} \Bigg({x_0^2  \bigg( \tilde{\tilde
A}_1(x_0)-\tilde{\tilde A}_2(x_0)+\tilde{\tilde
A}_3(x_0)+\tilde{\tilde A}_4(x_0)-\tilde{\tilde
A}_5(x_0)+\tilde{\tilde A}_6(x_0)\bigg)\over m_0^2-q^2+x_0^2M^2}
\Bigg)\bigg], \nonumber
\end{eqnarray}
with
\begin{eqnarray}
x_0={\sqrt{(Q^2+s_0-M^2)^2+4M^2(Q^2+m_b^2)}-(Q^2+s_0-M^2)\over 2M^2}
.
\end{eqnarray}

It can be easily observed that the sum rules for form factors
$F_1(q^2)$ and $F_2(q^2)$ are the same as that for $f_1(q^2)$ and
$f_2(q^2)$ respectively, namely, $F_1(q^2)=f_1(q^2), \,\,\,
F_2(q^2)=f_2(q^2)$. In addition, we indeed  find that form factors
$f_2(q^2)$ and $F_2(q^2)$ are exactly equal to zero at  zero
momentum transfer, which indicates that they  do not play any role
to the radiative decay  $\Lambda_b \to \Lambda + \gamma$.

Now we are going to investigate the asymptotic behavior of form
factors $f_1(q^2)$ and $f_2(q^2)$ in the limit of $Q^2 \to \infty$.
The form factors are dominated by the configurations with the
recoiled $s$ quark taking most  of  the momentum of $\Lambda$
baryon. Employing the asymptotic distribution amplitudes of
$\Lambda$ baryon and expanding the sum rules given in Eqs. (\ref{sum
relus for f1}) and (\ref{sum relus for f2}) in $1/Q^2$, we can get
the following expressions of form factors in the limit of large
momentum transfer
\begin{eqnarray}
f_2(q^2)&=&{3 \over 2}{f_{\Lambda} \over f_{\Lambda_b}}{ M^2 \over
M_B^2} {1 \over Q^6}(1+ {\lambda_1 \over f{\Lambda}})\bigg(s_0^2
M_B^2 e^{-(s_0-m_{\Lambda_b}^2)/M_B^2} + \int_{m_b^2}^{s_0}d s \,
s^2 e^{-(s-m_{\Lambda_b}^2)/M_B^2} \bigg),  \\
f_1(q^2)&=&{1 \over 6}{f_{\Lambda} \over f_{\Lambda_b}} { M \over
M_B^2} {1 \over Q^6} \left[(12-{ M^2 \over M_B^2})-3{\lambda_1 \over
f_{\Lambda}}(4-{ M^2 \over M_B^2})\right] \bigg(s_0^3 M_B^2
e^{-(s_0-m_{\Lambda_b}^2)/M_B^2} +\int_{m_b^2}^{s_0}d s \, s^3
e^{-(s-m_{\Lambda_b}^2)/M_B^2} \bigg), \nonumber
\end{eqnarray}
from which one can observe that $f_2(q^2)$ is suppressed by  $1
/Q^2$ compared to the expected asymptotic behavior $f_2(Q^2) \sim
1/Q^4$ in terms of the analysis on the electromagnetic pion form
factor \cite{asymptonic fom factor of pion 1,asymptonic fom factor
of pion 2}. It is mentioned in \cite{form factor of nucleon} that
the true asymptotic behavior of $f_2(q^2) \sim 1/Q^4$ can be
reproduced as a part of $O(\alpha_s^2)$ corrections, corresponding
to the hard scattering mechanism, in the framework of LCSR approach.
We also note that the pure contributions from the leading twist
distribution amplitudes of $\Lambda$ baryon will give rise to the
asymptotical behavior $f_2(Q^2) \sim 1/Q^8$, which is suppressed by
one more power of $1 /Q^2$ compared to that from higher twist
distribution amplitudes. This observation also indicates that the
soft form factors of $\Lambda_b \to \Lambda$ are dominated by the
distribution amplitudes with ``wrong helicity " instead of the
leading twist ones, which is in agreement with the analysis of soft
contributions for nucleon form factors \cite{form factor of nucleon}
and $\Lambda_b \to p \, l \bar{\nu}$ transition form factors
\cite{m.q. huang}.

\subsection{Light-cone sum rules for $\Lambda_b \to \Lambda + l^{+}
l^{-}$}

  Different from the radiative
decay $\Lambda_b \to \Lambda \gamma$, both the matrix elements given
by the  tensor operator $O_{7 \gamma}$ and vector-like currents
$O_{9}$, $O_{10}$ sandwiching between the $\Lambda_b $ and $\Lambda$
states can contribute to the decay $\Lambda_b \to \Lambda + l^{+}
l^{-}$. As the former matrix elements can be directly borrowed from
the radiative decay, we only need to deal with latter one within the
LCSR approach.  We start with the following correlation function
\begin{eqnarray}
z^{\nu} \tilde{T}_{\nu}(P,q)=i z^{\nu} \int d^4 x e^{-i q \cdot x}
\langle 0 |T \{ j_{\Lambda_b}(0) \tilde{j}_{\nu}(x)\} | \Lambda  (P)
\rangle, \label{correlation function V-A}
\end{eqnarray}
where the current $\tilde{j}_{\nu}(x)$ is given by
\begin{eqnarray}
\tilde{j}_{\nu}(x)=  \bar{b}(x) \gamma_{\nu}(1-\gamma_5)s(x)\,\,.
\end{eqnarray}
We can write the phenomenological representation of the correlator
at the hadronic level simply  as
\begin{eqnarray}
z^{\nu} \tilde{T}_{\nu}=2 f_{\Lambda_b} {(z \cdot P^{\prime})^2
\over m_{\Lambda_b}^2-{P^{\prime}}^2} \bigg[g_1 \not \! z -g_2 \not
\! z \not \! q -G_1 \not \! z \gamma_5  -G_2 \not \! z \not \! q
\gamma_5 \bigg] \Lambda(P)+...., \label{correlation function at
hadron level V-A current}
\end{eqnarray}
where the contributions from $g_3$ and $G_3$ are proportional to the
higher power of $1/P$ in the infinite momentum kinematics $P\sim
\infty, \,\, q \sim const, \,\, z \sim 1/P$  and hence are omitted
in this paper. On the other hand,   the correlation function at the
quark level can be calculated in the framework of perturbative
theory to the leading order of $\alpha_s$ as
\begin{eqnarray}
z^{\nu} T_{\nu}&&=-2(z \cdot P)^2 \not \! z (1-\gamma_5)\Lambda(P)
\bigg \{ \int {\cal{D}} x {x_3 A_1
(x_i) \over (x_3P+q)^2-m_b^2} \nonumber \\
&& \hspace{1.0 cm}+M^2 \int_0^1 dx_3 \,
x_3^2{-2\tilde{A}_1(x_3)+\tilde{A}_2(x_3)-\tilde{A}_3(x_3)
-\tilde{A}_4(x_3)+\tilde{A}_5(x_3)\over [(x_3P+q)^2-m_b^2]^2}\nonumber \\
&& \hspace{1.0 cm}+2 M^4 \int_0^1 dx_3 \, x_3^3{\tilde{\tilde
A}_1(x_3)-\tilde{\tilde A}_2(x_3)+\tilde{\tilde
A}_3(x_3)+\tilde{\tilde A}_4(x_3)-\tilde{\tilde
A}_5(x_3)+\tilde{\tilde A}_6(x_3)\over [(x_3P+q)^2-m_b^2]^3}\bigg\}\nonumber \\
&& +2  (z \cdot P)^2  \not \! z \not \! q
(1+\gamma_5)\Lambda(P)\bigg\{ M \int_0^1 dx_3 \, x_3
{-\tilde{A}_1(x_3)+\tilde{A}_2(x_3)-\tilde{A}_3(x_3) \over
[(x_3P+q)^2-m_b^2]^2 } \nonumber \\
&&  +2 M^3 \int_0^1 dx_3 \, x_3^2{\tilde{\tilde
A}_1(x_3)-\tilde{\tilde A}_2(x_3)+\tilde{\tilde
A}_3(x_3)+\tilde{\tilde A}_4(x_3)-\tilde{\tilde
A}_5(x_3)+\tilde{\tilde A}_6(x_3)\over [(x_3P+q)^2-m_b^2]^3}
\bigg\}+.... \label{correlation function V-A theoretically}
\end{eqnarray}
Matching the correlation function obtained in the two different
representations and performing the Borel transformation with respect
to the variable $P^2$, we can achieve the sum rules for the form
factors $g_1$ and $g_2$ responsible for  $\Lambda_b \to \Lambda +
l^{+} l^{-}$ decay as
\begin{eqnarray}
f_{\Lambda_b} g_1(q^2) e^{-m_{\Lambda_b}^2 / M_B^2}&=& \int_{x_0}^1
dx_3 e^{-s/M_B^2} \bigg[ \bigg( \int_0^{1-x_3}dx_3
A_1(x_1,1-x_1-x_3,x_3)\bigg)\nonumber \\
&& \hspace{1.0 cm}-{M^2 \over
M_B^2}\bigg(-2\tilde{A}_1(x_3)+\tilde{A}_2(x_3)-\tilde{A}_3(x_3)
-\tilde{A}_4(x_3)+\tilde{A}_5(x_3)\bigg)\nonumber \\
&& \hspace{1.0 cm}+{M^4 \over M_B^4}\bigg( \tilde{\tilde
A}_1(x_3)-\tilde{\tilde A}_2(x_3)+\tilde{\tilde
A}_3(x_3)+\tilde{\tilde A}_4(x_3)-\tilde{\tilde
A}_5(x_3)+\tilde{\tilde A}_6(x_3)\bigg)\bigg] \nonumber \\
&& -{M^2 x_0^2 e^{-s_0/M_B^2}\over m_b^2-q^2+x_0^2M^2} \bigg[ \bigg(
-2\tilde{A}_1(x_0)+\tilde{A}_2(x_0)-\tilde{A}_3(x_0)
-\tilde{A}_4(x_0)+\tilde{A}_5(x_0)\bigg)\nonumber \\
&&\hspace{1.0 cm}-{M^2 \over M_B^2} \bigg( \tilde{\tilde
A}_1(x_0)-\tilde{\tilde A}_2(x_0)+\tilde{\tilde
A}_3(x_0)+\tilde{\tilde A}_4(x_0)-\tilde{\tilde
A}_5(x_0)+\tilde{\tilde A}_6(x_0)\bigg)\label{sum relus for g1} \\
&& +M^2 {d \over dx_0} \Bigg({x_0^2  \bigg( \tilde{\tilde
A}_1(x_0)-\tilde{\tilde A}_2(x_0)+\tilde{\tilde
A}_3(x_0)+\tilde{\tilde A}_4(x_0)-\tilde{\tilde
A}_5(x_0)+\tilde{\tilde A}_6(x_0)\bigg)\over m_0^2-q^2+x_0^2M^2}
\Bigg)\bigg],  \nonumber
\end{eqnarray}
and
\begin{eqnarray}
f_{\Lambda_b} g_2(q^2) e^{-m_{\Lambda_b}^2 / M_B^2}&=&  -M\bigg\{
\int_{x_0}^1 {dx_3 \over x_3}e^{-s/M_B^2} \bigg[ {1 \over M_B^2}
\bigg(-\tilde{A}_1(x_3)+\tilde{A}_2(x_3)-\tilde{A}_3(x_3)\bigg)\nonumber \\
&& \hspace{1.0 cm}-{M^2 \over M_B^4}\bigg(\tilde{\tilde
A}_1(x_3)-\tilde{\tilde A}_2(x_3)+\tilde{\tilde
A}_3(x_3)+\tilde{\tilde A}_4(x_3)-\tilde{\tilde
A}_5(x_3)+\tilde{\tilde A}_6(x_3)\bigg)\bigg] \nonumber \\
&& +{x_0 e^{-s_0/M_B^2}\over m_b^2-q^2+x_0^2M^2} \bigg[ \bigg(
-\tilde{A}_1(x_3)+\tilde{A}_2(x_3)-\tilde{A}_3(x_3)\bigg)\nonumber \\
&&\hspace{1.0 cm}-{M^2 \over M_B^2} \bigg( \tilde{\tilde
A}_1(x_0)-\tilde{\tilde A}_2(x_0)+\tilde{\tilde
A}_3(x_0)+\tilde{\tilde A}_4(x_0)-\tilde{\tilde
A}_5(x_0)+\tilde{\tilde A}_6(x_0)\bigg) \label{sum relus for g2}\\
&& +M^2 x_0 {d \over dx_0} \Bigg({x_0  \bigg( \tilde{\tilde
A}_1(x_0)-\tilde{\tilde A}_2(x_0)+\tilde{\tilde
A}_3(x_0)+\tilde{\tilde A}_4(x_0)-\tilde{\tilde
A}_5(x_0)+\tilde{\tilde A}_6(x_0)\bigg)\over m_0^2-q^2+x_0^2M^2}
\Bigg)\bigg]\bigg\}\,\,.\nonumber
\end{eqnarray}
In addition,  the sum rules for form factors satisfy
$G_1(q^2)=g_1(q^2)$, $G_2(q^2)=g_2(q^2)$ to the accuracy considered
in this work. These relations will be broken down with the inclusion
of higher power corrections in $1/m_b$  and higher order corrections
in $\alpha_s$. Furthermore, we can also get the following relations
between form factors $f_i$ and $g_i (i=1,2)$ by comparing the
revelent sum rules
\begin{eqnarray}
 f_1(q^2)=q^2 g_2(q^2), \hspace{1 cm}  f_2(q^2)=g_1(q^2),
\end{eqnarray}
which are consistent with those derived in \cite{c.q. geng 4, Aliev
1,Hiller:2001zj} based on the analysis of heavy quark symmetry and
the large energy symmetry.


\section{Numerical analysis of sum rules for form factors}

Now we are going to calculate form factors $g_2(q^2)$  and
$f_2(q^2)$ revelent to the $\Lambda_b \to \Lambda + \gamma$ and
$\Lambda_b \to \Lambda + l^{+} l^{-}$ decays numerically. Firstly,
we collect the input parameters used in this paper as below
\cite{PDG, m.q. huang lambda_b to proton, ioffe, bauer}
\begin{equation}
\begin{array}{ll}
G_F=  1.166 \times 10^{-2} {\rm{GeV}^{-2}}, &
|V_{ts}|=41.61^{+0.10}_{-0.80} \times 10^{-3},
\\
|V_{tb}|=0.9991,  & m_b=(4.68 \pm 0.03) {\rm{GeV}},
\\
m_c(m_c) = 1.275 ^{+0.015}_{-0.015} {\rm{GeV}}, &m_s(1
{\rm{GeV}})=(142 \pm 28) {\rm{MeV}},
\\
m_{\Lambda_b}=5.62 {\rm{GeV}}, &  m_{\Lambda}=1.12 {\rm{GeV}},
\\
f_{\Lambda_b}=3.9 ^{+0.4}_{-0.2}\times 10^{-3} {\rm{GeV}}^{2}, &
f_{\Lambda}=6.0^{+0.4}_{-0.4} \times 10^{-3} {\rm{GeV}}^{2}
\\
\lambda_1=-1.3^{+0.2}_{-0.2} \times 10^{-2} {\rm{GeV}}^{2}, \,\,\, &
s_0=39 \pm 1 {\rm{GeV}}^2.\label{inputs}
\end{array}
\end{equation}
It should be noted that the normalization constants of LCDAs for
$\Lambda_b$ and $\Lambda$ baryons, namely $f_{\Lambda_b}$,
$f_{\Lambda}$ and $\lambda_1$, are all evaluated at the scale $\mu=
1 \mbox {GeV}$. As for the choice of the threshold parameter $s_0$,
one should determine it by demanding the sum rules results to be
relatively stable in allowed regions for Borel mass $M_B^2$, the
value of which should be around the mass square of the corresponding
first excited states.

With   all the parameters, we can proceed to compute the numerical
values of the form factors. In principle, the form factors
$g_2(q^2)$  and $f_2(q^2)$ should not depend on the Borel masse
$M_B^2$ in a complete theory. However, as we truncate the operator
product expansion up to the leading conformal spin of distribution
amplitudes for $\Lambda$ baryon in the leading Fock configuration
and keep the perturbative expansion in $\alpha_s$ to leading order,
a manifest dependence of the form factors on the Borel parameter
$M_B^2$  would emerge  in practice. Therefore, one should look for a
working ``window", where the results only mildly vary with respect
to the Borel mass, so that the truncation is reasonable and
acceptable.

\begin{figure}[tb]
\begin{center}
\begin{tabular}{ccc}
\includegraphics[scale=0.6]{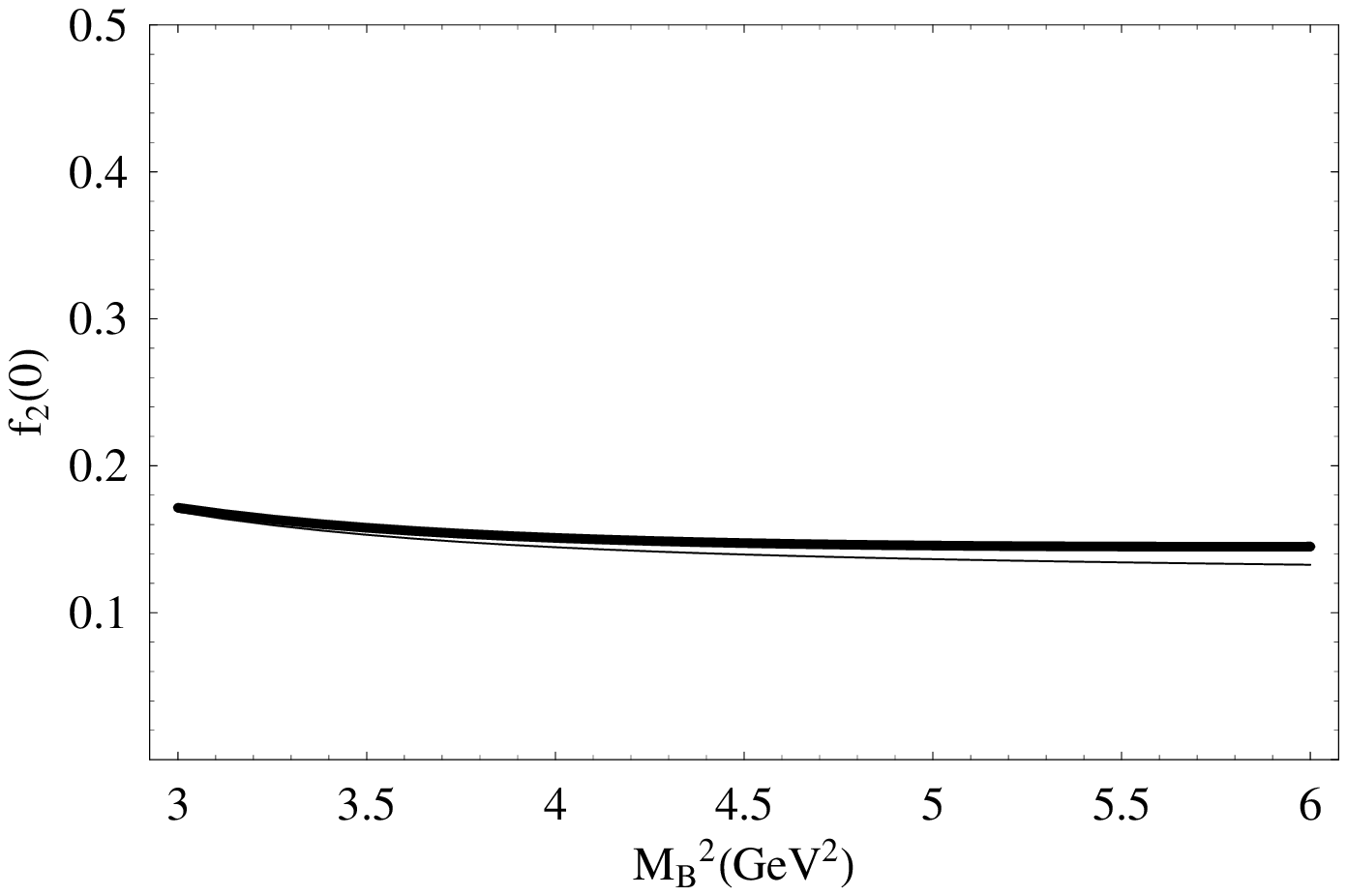}
\includegraphics[scale=0.6]{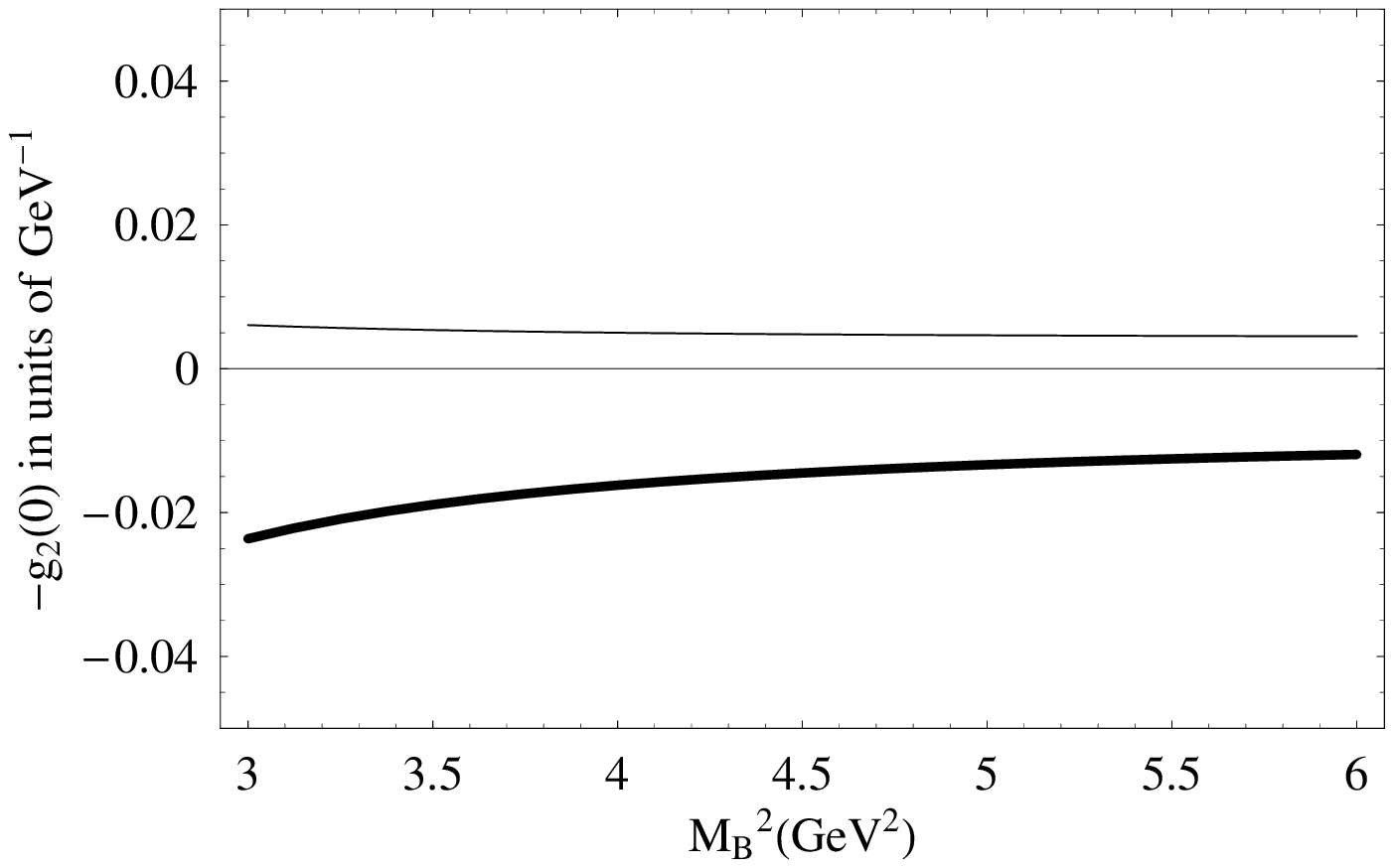}
 \vspace{-1cm}
\end{tabular} \caption{Dependence of
form factors $f_2(0)$ and $g_2(0)$ 
on the working Borel window $M_B^2 \in [3.0,6.0] {\mbox { GeV}^2}$
with the chosen threshold parameter $s_{\Lambda_b}^0=39
{\rm{GeV}}^2$. Here the bold line denotes the contributions from the
distribution amplitudes up to twist-6, while the thin line
represents the contributions from leading twist ones only.
}\label{form factor f2 and g2}
\end{center}
\end{figure}

 Firstly,  we concentrate on  the form
factors at zero momentum transfer. For the form factor $f_2(0)$, we
require that the contributions from the higher resonances and
continuum states hold the fraction less than 25 \% in the total sum
rules and the value of $f_2(0)$ does not vary drastically within the
selected region for the Borel mass.  In view of these
considerations, the Borel parameter $M_B^2$ should not be too large
in order to insure that the contributions from the higher states are
exponentially damped as can be observed form Eqs. (\ref{sum relus
for f1}), (\ref{sum relus for f2}), (\ref{sum relus for g1}) and
(\ref{sum relus for g2}) and the global quark-hadron duality is
satisfactory. On the other hand, the  Borel mass   could not be too
small for the  validity of OPE near the light-cone for the
correlation function in deep Euclidean region, since the
contributions of higher twist distribution amplitudes amount the
higher order of ${1 / M_B^2}$ to the perturbative part. With the
chosen threshold value $s_0=39 {\rm{GeV}}^2$, we   indeed find the
Borel platform $M_B^2 \in [3.0, 6.0] \mathrm{GeV}^2$, which is
plotted in Fig. \ref{form factor f2 and g2}. The number of $f_2(0)$
is $0.15^{+0.02}_{-0.02}$, where we have combined the uncertainties
from the variation of Borel parameters, fluctuation of threshold
value, errors of $b$ quark mass and uncertainties from the
non-perturbative parameters in the distribution amplitudes of
$\Lambda$ baryon together. The errors on the form factor $f_2(0)$
are estimated within the level of 20 \% as expected by the general
understanding of the theoretical framework. Following the same
method, we can continue to estimate the numerical results for the
form factor $g_2(0)$ within the selected Borel window as displayed
in Fig. \ref{form factor f2 and g2}.  The contributions from the
excited resonance and continuum state are required to be less than
10 \%    here.   It can be observed that $g_2(0)= 1.3^{+0.2}_{-0.4}
\times 10^{-2}\, \mbox {GeV}^{-1}$ with the given Borel window
$M_B^2 \in [3.0, 6.0]$ {GeV}$^2$.

Next, we can further investigate the $q^2$ dependence of the form
factors $f_2$ and $g_2$ based on the sum rules given in Eqs.
(\ref{sum relus for f2}) and (\ref{sum relus for g2}). The OPE for
the correlation function in Eqs. (\ref{correlation function tensor
current}) and (\ref{correlation function V-A}) near the light-cone
is valid only at small or intermediate squared momentum transfer,
$0<q^2<m_b^2-2m_b\Lambda_{QCD}$, which   ensure the good stability
of the sum rules in Eqs. (\ref{sum relus for f1}), (\ref{sum relus
for f2}), (\ref{sum relus for g1}) and (\ref{sum relus for g2}) with
the variation of $M_B^2$.   The reason is that the light-cone
expansion is expected to break down when $q^2$ is approaching
$m_b^2$ and hence contributions from the higher twist distribution
amplitudes increase rapidly. In phenomenology, we extend the  form
factors to large squared  momentum transfer   by the  double-pole
model
\begin{eqnarray}
\label{form} \xi_{i}(q^2)={\xi_i(0) \over 1-a_1
q^2/m_{\Lambda_b}^{2}+a_2 q^4/m_{\Lambda_b}^{4}}, \label{pole model
of form factors}
\end{eqnarray}
  where $\xi_i$ denotes the
form factors $f_2$ and $g_2$. The  parameters $a_1$ and $a_2$ can be
fixed by the matching of form factors corresponding to the small and
intermediate momentum transfer calculated in the LCSR approach. Our
results   are  grouped in Table~\ref{di-fit}, where the values of
form factors in terms of  the leading twist sum rules (twist-3) are
also presented for a compassion.

\begin{table}[htb]
\caption{Numerical results for the form factors $f_2(0)$, $g_2(0)$
and  parameters $a_1$ and $a_2$ involved in the double-pole fit of
eq. (\ref{pole model of form factors}) for both twist-3 and twist-6
sum rules with $M_B^2 \in [3.0, 6.0]~\mbox{GeV}^2$, $s_0=39 \pm
1~\mbox{GeV}^2$,
 together with results from COZ \cite{COZ DAs} and FZOZ \cite{FZOZ DAs}
models for distribution amplitudes of $\Lambda$ baryon in the
leading twist sum rules and results obtained in QCDSR \cite{Huang}.
} \label{di-fit}
\begin{tabular}{cccccc}
\hline \hline
parameter  & {COZ}&  {FZOZ} &QCDSR & twist-3  & up to twist-6\\
\hline $f_2(0)$ & $0.74^{+0.06}_{-0.06}$ & $0.87^{+0.07}_{-0.07}$ &
$0.45$ &$0.14^{+0.02}_{-0.01}$
&$0.15^{+0.02}_{-0.02}$\\
{$a_1$} & $2.01^{+0.17}_{-0.10}$  & $2.08^{+0.15}_{-0.09}$ &0.57
&$2.91^{+0.10}_{-0.07}$
&$2.94^{+0.11}_{-0.06}$\\
{$a_2$}& $1.32^{+0.14}_{-0.08}$ &$1.41^{+0.11}_{-0.08}$&
$-0.18$&$2.26^{+0.13}_{-0.08}$
&$2.31^{+0.14}_{-0.10}$ \\
\hline
 $g_2 (0) (10^{-2}\rm{GeV^{-1}})$  & $-2.4^{+0.3}_{-0.2}
 $ & $-2.8^{+0.4}_{-0.2}  $ & $-1.4$ &$-0.47^{+0.06}_{-0.06}
 $ &$1.3^{+0.2}_{-0.4} $
\\
 {$a_1$}& $2.76^{+0.16}_{-0.13}$ & $2.80^{+0.16}_{-0.11}$ &2.16 &$3.40^{+0.06}_{-0.05}$
 &$2.91^{+0.12}_{-0.09}$\\
{$a_2$} &$2.05^{+0.23}_{-0.13}$ &$2.12^{+0.21}_{-0.13}$ &1.46
&$2.98^{+0.09}_{-0.08}$
&$2.24^{+0.17}_{-0.13}$ \\
\hline \hline
\end{tabular}
\end{table}

As can be observed from Table \ref{di-fit}, the form factor $f_2$ is
dominated by the contributions from the leading twist distribution
amplitudes for $\Lambda$ baryon, and the corrections owing to the
higher twist distribution amplitudes  are less than 10 \%. In
contrast, the higher twist distribution amplitudes play an important
role in the form factor $g_2$, where the numbers predicted in the
twist-3 LCSR even differs from  that of the whole sum rules in the
sign. This observation is in agreement with the studies on
semi-leptonic decay $\Lambda_c \to \Lambda l^{+} l$ \cite{m.q.
huang}  and nucleon form factor \cite{form factor of nucleon}.
Predictions on these form factors can be systematically improved by
including the higher conformal partial waves  and higher order
perturbative corrections.

Apart from the general description of distribution amplitudes based
on the conformal symmetry of QCD Lagrangian, there exist two
concrete models taking into account the first a few conformal
partial waves for the $\Lambda$ baryons,   COZ model \cite{COZ DAs},
and FZOZ model \cite{FZOZ DAs}.   The   expressions of twist-3
distribution amplitudes for COZ model $A_1^{COZ}(x_i)$ and FZOZ
model $A_1^{FZOZ}(x_i)$ can be written as
\begin{eqnarray}
\label{COZ DAs} A_1^{COZ}(x_1,x_2,x_3)&=&
-42\phi_{as}(x_1,x_2,x_3)[0.26(x_3^2+x_2^2)+0.34x_1^2-0.56x_2x_3-0.24x_1(x_2+x_3)],
\\A_1^{FZOZ}(x_1,x_2,x_3)&=&
-42\phi_{as}(x_1,x_2,x_3)[0.093(x_3^2+x_2^2)+0.376x_1^2-0.194x_2x_3-0.207x_1(x_2+x_3)],
 \nonumber\\
\phi_{as}(x_1,x_2,x_3) &=& 120x_1x_2x_3.
\end{eqnarray}
Substituting the above distribution amplitudes for $\Lambda$ baryon
into sum rules  Eqs. (\ref{sum relus for f2}) and (\ref{sum relus
for g2}) and repeating the same procedure, we can get   the
parameters for  $a_1$ and $a_2$ accounting for the form factor $f_2$
and $g_2$ as grouped in Table~\ref{di-fit}. From the table, we can
see that  soft contributions to the transition form factors $f_2$
and $g_2$ in the COZ and FZOZ models are approximately five times
larger than that for the LCDAs of $\Lambda$ baryon based on the
conformal spin expansion.  This is similar to that observed in the
studies of pion form factor \cite{asymptonic fom factor of pion 1},
nucleon form factor \cite{form factor of nucleon} and also
$\Lambda_c \to \Lambda$ transition \cite{m.q. huang}.

The form factors are also calculated in the QCDSR \cite{Huang} using
the heavy quark symmetry to reduce the number of independent form
factors. We have translated their results with the help of Eq.
(\ref{relatins of from factors in HQET 1}-\ref{relatins of from
factors in HQET 2}).
It can be observed from Table \ref{di-fit} that the numbers of
$g_2(0)$ obtained in QCDSR differ from that extracted from twist-6
sum rules on the light-cone in the sign. Besides, $f_2(0)$ in the
framework of QCDSR is about three times larger than that given by
LCSR approach presented here. In particular, the $q^2$ dependence of
the from factor $f_2$ between these two methods are quite different.
It grows slowly with the increase of squared momentum transfer $q^2$
in QCDSR to reach 0.83 at the maximal momentum transfer. But it
rises drastically in our approach with the increase of $q^2$ to
reach 2.3 at $q^2_{max}=(m_{\Lambda_b}-m_{\Lambda})^2=20.3 {\mbox
{GeV}}^{2}$. Such differences between form factors  will lead to
quite different predictions on values of decay width and
forward-backward asymmetry for semi-leptonic decay $\Lambda_b \to
\Lambda l^{+} l^{-}$.

The hard contributions to the form factor $f_2(0)$ involving two
hard gluons exchange have  also been investigated \cite{HLLW} to the
leading twist distribution amplitudes for $\Lambda$ baryon in the
framework of PQCD approach. The values of $f_2(0)$ is computed as
$(1.2 \sim 1.6) \times 10^{-2}$, which is about one order smaller
than   the form factor presented here in terms of the LCSR approach.
As a matter of fact, large soft corrections have been observed in
the non-leptonic charmed meson decays in the perturbative QCD
approach based on $k_T$ factorization \cite{soft effect in PQCD}.
Therefore, it seems that the form factors responsible for $\Lambda_b
\to \Lambda$ transition may be dominated by the soft gluons exchange
between valence quarks inside the $\Lambda_b$ and $\Lambda$ baryons.

\section{Phenomenology of $\Lambda_b \to \Lambda + \gamma$
and $\Lambda_b \to \Lambda + l^{+} l^{-}$}

Utilizing the  form factors derived above, we can now proceed to
perform the calculations of decay rate, polarization asymmetry and
forward-backward asymmetry.

\subsection{Decay width and $\Lambda$ polarization asymmetry of $\Lambda_b \to \Lambda + \gamma$}
\label{numerical analysis of radiative decay}

In this subsection, we present the formulae of decay rate and
polarization asymmetry of $\Lambda$ baryon for $\Lambda_b \to
\Lambda + \gamma$, the latter of which can be used to analyze the
helicity structures of effective Lagrangian at the quark level. The
four-spin vector $s^{\mu}$ of $\Lambda$ baryon can be  defined in
its rest frame as
\begin{eqnarray}
(s^{\mu})_{r.s.}=(0, \,\, \hat{\bf{\xi}}), \label{spin vector of
lambda baryon 1}
\end{eqnarray}
which can be directly transformed into the rest frame of $\Lambda_b$
baryon by the Lorentz boost
\begin{eqnarray}
s^{\mu}=({{\bf{p}}_{\Lambda}\cdot {\hat{\bf{\xi}}} \over
m_{\Lambda}}, \,\, {\bf{\hat{\xi}}}+ {s_0 \over
E_{\Lambda}+m_{\Lambda}} {\bf{p}}_{\Lambda}), \label{spin vector of
lambda baryon 2}
\end{eqnarray}
with ${\bf{p}}_{\Lambda}$ and $E_{\Lambda}$ being the three-momentum
and energy of $\Lambda$ baryon. Then, the following relations can be
directly read
\begin{eqnarray}
v \cdot s= {1-x^2 \over 1+x^2} {\bf{\hat{p}}} \cdot {\bf{s}} =
{1-x^2 \over 2x} {\bf{\hat{p}}} \cdot {\bf{\hat{\xi}}}, \label{spin
relation}
\end{eqnarray}
 where $x=m_{\Lambda}/m_{\Lambda_b}$, ${\bf{\hat{p}}}$ is a unite
vector along the  momentum of $\Lambda$ baryon    and
$v={p_{\Lambda_b}} / m_{\Lambda_b}$ is the four-velocity vector of
$\Lambda_b$.

Making use of Eqs. (\ref{parameterization of tensor current 1}) and
(\ref{parameterization of tensor current 2}), the decay width of
unpolarized $\Lambda_b$ into $\Lambda$ with a definite polarization
vector $s$ can be derived as \cite{Mannel,Huang}
\begin{eqnarray}
\Gamma(\Lambda_b \to \Lambda \gamma) = {\alpha_{em} G_F^2 \over 64
m_{\Lambda_b}^3\pi^4}|V_{tb}|^2|V_{ts}|^2 |C_7^{eff}|^2 (1-x^2)^3
(m_b^2+m_s^2)[f_2(0)]^2 \bigg[1+{2 x \over 1-x^2}{m_b^2-m_s^2 \over
m_b^2+m_s^2 } (v \cdot s)\bigg]. \label{decay width of radiative
decay}
\end{eqnarray}
  For the convenience of
comparing with the experimental data, we can rewrite the  Eq.
(\ref{decay width of radiative decay}), in the standard form
\cite{PDG} making use  of Eq. (\ref{spin relation}) as
\begin{eqnarray}
\Gamma(\Lambda_b \to \Lambda \gamma) = {1 \over 2}\Gamma_0 [1+\alpha
\,\, {\bf{\hat{p}}} \cdot {\bf{s}} ]= {1 \over 2}\Gamma_0
[1+\alpha^{\prime} \,\, {\bf{\hat{p}}} \cdot {\bf{\hat{\xi}}} ],
\label{form of polarization asymmetry}
\end{eqnarray}
with
\begin{eqnarray}
\Gamma_0 = {\alpha_{em} G_F^2 \over 32
m_{\Lambda_b}^3\pi^4}|V_{tb}|^2|V_{ts}|^2 |C_7^{eff}|^2 (1-x^2)^3
(m_b^2+m_s^2)[f_2(0)]^2 , \label{total decay width of radiative
decay}
\end{eqnarray}
and
\begin{eqnarray}
\alpha={2x \over 1+x^2}\alpha^{\prime}={2x \over 1+x^2}{m_b^2-m_s^2
\over m_b^2+m_s^2 }. \label{polarization asymmetry of radiative
decay}
\end{eqnarray}
As can be observed from Eq. (\ref{polarization asymmetry of
radiative decay}), the polarization parameters $\alpha $  and
$\alpha^{\prime}$ are free of the pollution due to strong
interactions \cite{x.g. he,Huang} and only depend on the relative
strength of left- and right- handed couplings between quarks.
Utilizing the inputs given in Eq. (\ref{inputs}), we can get the
polarization parameters as $\alpha=0.381 \pm 0.001, \,\,
\alpha^{\prime}=0.998 \pm 0.001$. Any distinct deviations from these
values would indicate the new physics beyond the SM.

Substituting the form factor $f_2(0)$    calculated in the last
section into Eq. (\ref{total decay width of radiative decay}), we
can achieve the decay rate  of radiative decay  $\Lambda_b \to
\Lambda + \gamma$ as shown in the Table~\ref{results of radiative
decay rate in the literature}.  Predictions on the branching
ratio(BR) of $\Lambda_b\to \Lambda \gamma$ can vary even by the
order of magnitude adopting different   models of distribution
amplitudes for $\Lambda$ baryon. We also collect the results of
decay rate for $\Lambda_b \to \Lambda + \gamma$ computed in other
approaches in Table~\ref{results of radiative decay rate in the
literature}, from which we can find that different methods   give
quite different predictions. Unfortunately,  only the upper bound
$1.3 \times 10^{-3}$ for BR of $\Lambda_b\to \Lambda \gamma$ decay
is available in experiment at present, so we have to wait for more
experimental data to discriminate existing models.


\begin{table}
\caption{Decay branching ratios (BR) of $\Lambda_b\to \Lambda
\gamma$   calculated in the LCSR approach with distribution
amplitudes of $\Lambda$ baryon in terms of the conformal spin
expansion with only  twist-3 and up to twist-6   together with
results from COZ and FZOZ model,  and results based on the form
factors from pole model \cite{Mannel}, covariant oscillator quark
model \cite{Mohanta}, heavy quark effective theory \cite{Cheng}, MIT
bag model \cite{Cheng}, non-relativistic quark model \cite{Cheng 2},
QCD sum rule approach \cite{Huang} and perturbative QCD approach
\cite{HLLW}, respectively.}
\begin{center}
\begin{tabular}{ccc c c c c c c}
  \hline
  \hline
  Model of DAs & twist-3 & up to twist-6 &  COZ & FZOZ \\
  \hline
  BR ($\times 10^{-5}$) & $0.63^{+0.17}_{-0.12} $& $0.73^{+0.15}_{-0.15} $ &
  $16^{+2}_{-4} $ & $22^{+5}_{-5}  $ \\
  \hline
  Model &PM &  COQM & HQET & BM
  & NRQM& QCDSR & PQCD \\
  \hline
  BR ($\times 10^{-5}$)& $1.0\sim 4.5$ & $0.23$ & $0.8\sim 1.5$
  & $0.4$ & 0.27 & $3.7\pm 0.5$& $ (0.0043 \sim 0.0086)  $ \\
  \hline
  \hline
\end{tabular}
\end{center}
\label{results of radiative decay rate in the literature}
\end{table}

\subsection{Decay width and dilepton distributions of $\Lambda_b \to \Lambda + l^{+} l^{-}$}

We adopt the transition form factors calculated in the sum rules up
to the twist-6 distribution amplitudes of $\Lambda$ baryon. The
long-distance effects originated from $c \bar{c}$ resonances on the
${\mbox {BR}}(\Lambda_b \to \Lambda l^+ l^-)$ are also discussed.
The differential decay width of $\Lambda_b \to \Lambda l^+ l^-$ in
the rest frame of $\Lambda_b$ baryon can be written as \cite{PDG},
\begin{equation}
{d\Gamma({\Lambda_b \to \Lambda l^+ l^-}) \over d q^2} ={1 \over (2
\pi)^3} {1 \over 32 m_{\Lambda_b}^3} \int_{u_{min}}^{u_{max}}
|{\widetilde{M}}_{\Lambda_b \to \Lambda l^+ l^-}|^2 du,
\label{differential decay width}
\end{equation}
where $u=(p_{\Lambda}+p_{l^{-}})^2$ and $q^2=(p_{l^+}+p_{l^-})^2$;
$p_{\Lambda}$, $p_{l^{+}}$ and $p_{l^{-}}$ are the four-momenta
vectors of $\Lambda$, $l^{+}$ and $l^{-}$ respectively.
${\widetilde{M}}_{\Lambda_b \to \Lambda l^+ l^-}$ is the  decay
amplitude after integrating over the angle between the $l^{-}$ and
$\Lambda$ baryon. The upper and lower limits of $u$ are given by
\begin{eqnarray}
u_{max}&=&(E^{\ast}_{\Lambda}+E^{\ast}_{l})^2-(\sqrt{E_{\Lambda}^{\ast
2}-m_{\Lambda}^2}-\sqrt{E_l^{\ast 2}-m_l^2})^2, \nonumber\\
u_{min}&=&(E^{\ast}_{\Lambda}+E^{\ast}_{l})^2-(\sqrt{E_{\Lambda}^{\ast
2}-m_{\Lambda}^2} +\sqrt{E_l^{\ast 2}-m_l^2})^2;
\end{eqnarray}
where $E^{\ast}_{\Lambda}$ and $E^{\ast}_{l}$ are the energies of
$\Lambda$ and $l^{-}$ in the rest frame of lepton pair
\begin{equation}
E^{\ast}_{\Lambda}= {m_{\Lambda_b}^2 -m_{\Lambda}^2 -q^2 \over 2
\sqrt{q^2}}, \hspace {1 cm} E^{\ast}_{l}={q^2\over 2\sqrt{q^2}}.
\end{equation}

\begin{figure}[tb]
\begin{center}
\begin{tabular}{ccc}
\vspace{-2cm}
\includegraphics[scale=0.6]{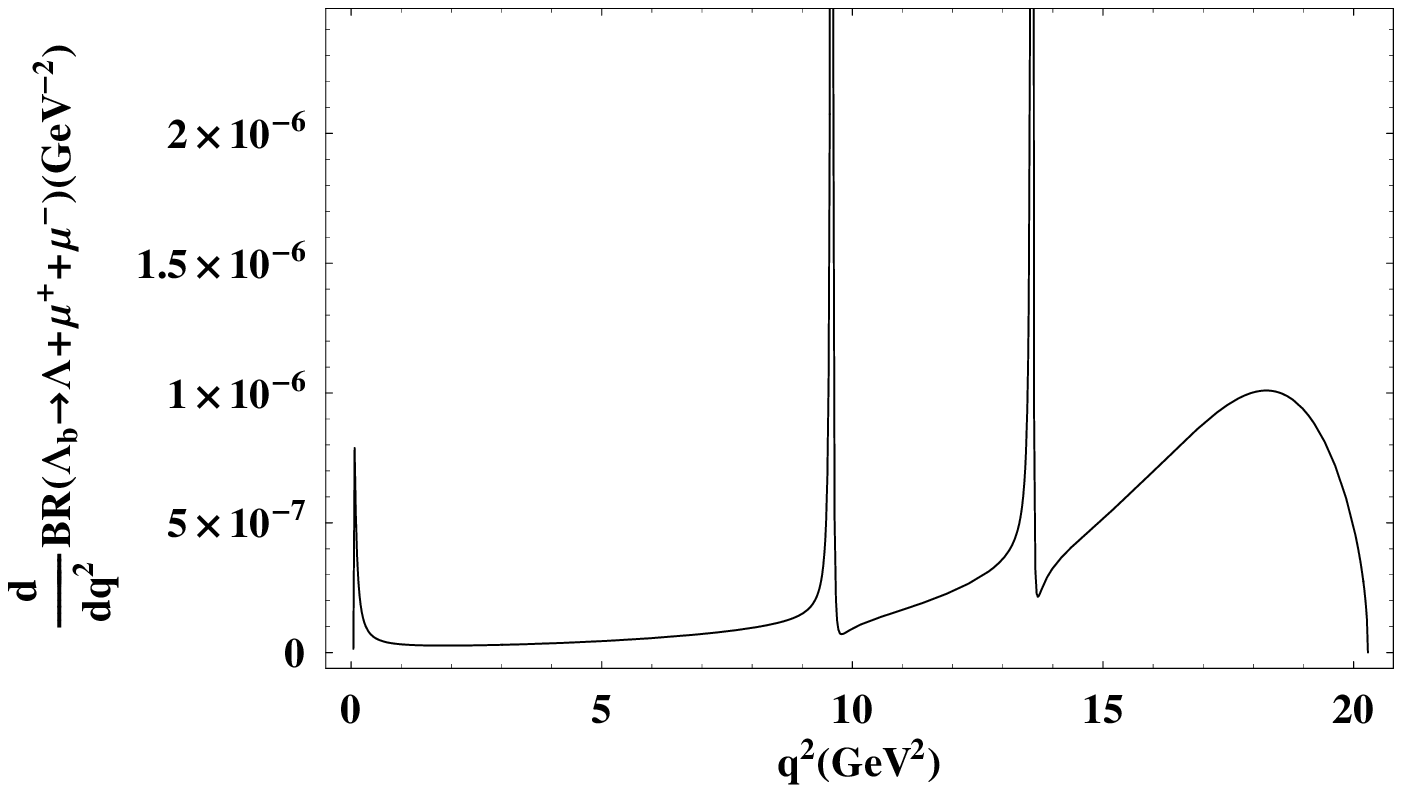}
\includegraphics[scale=0.6]{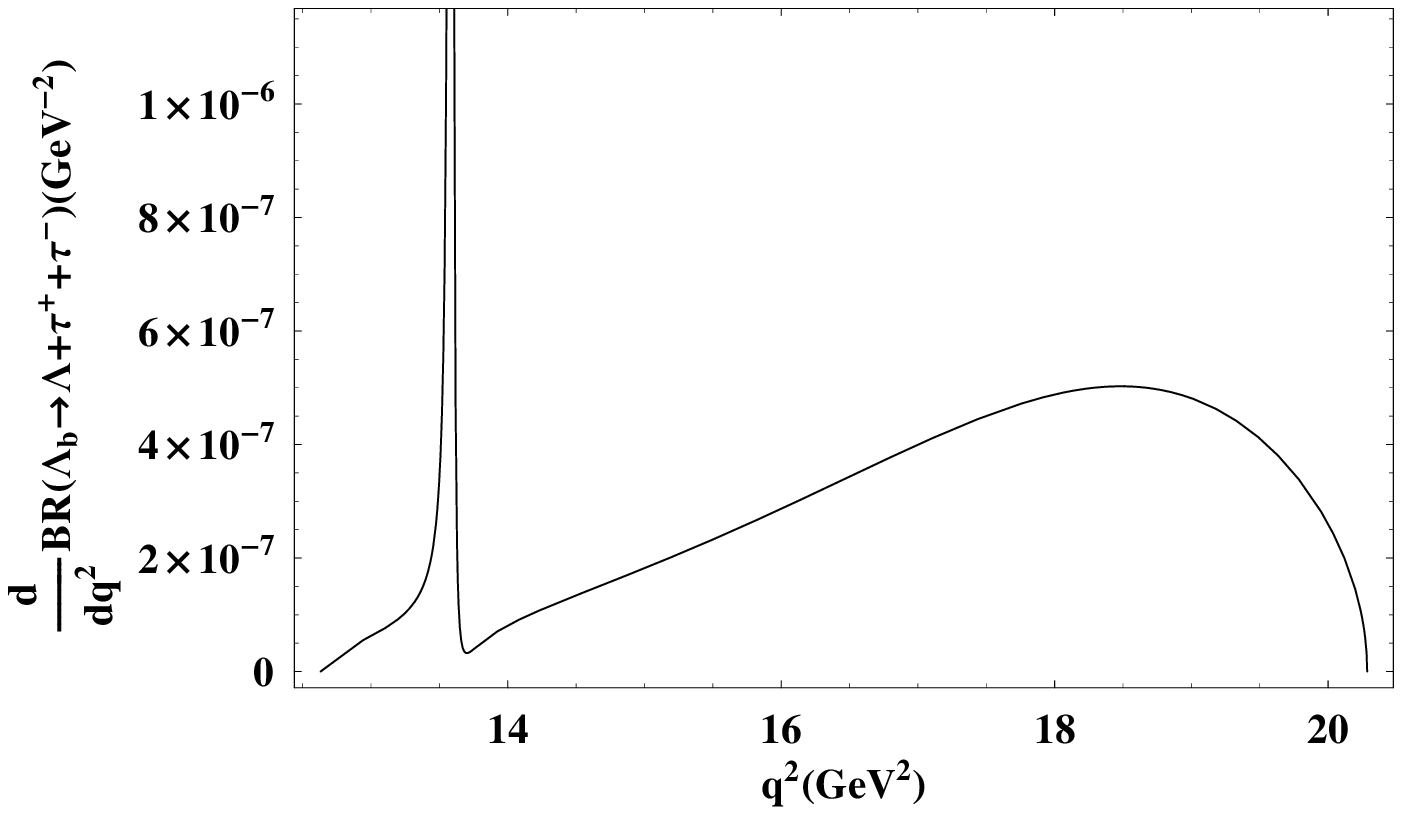}
\\
\includegraphics[scale=0.6]{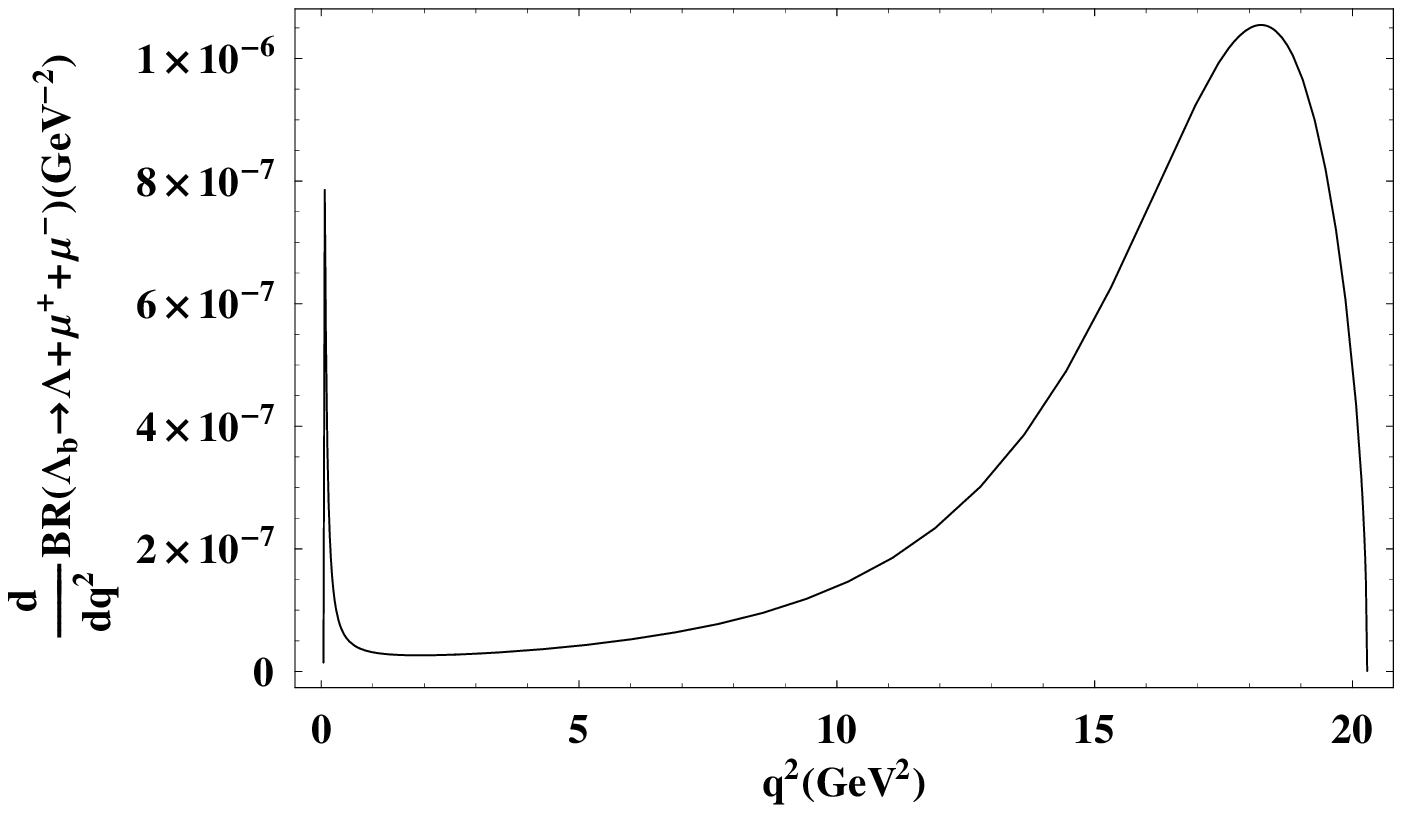}
\includegraphics[scale=0.6]{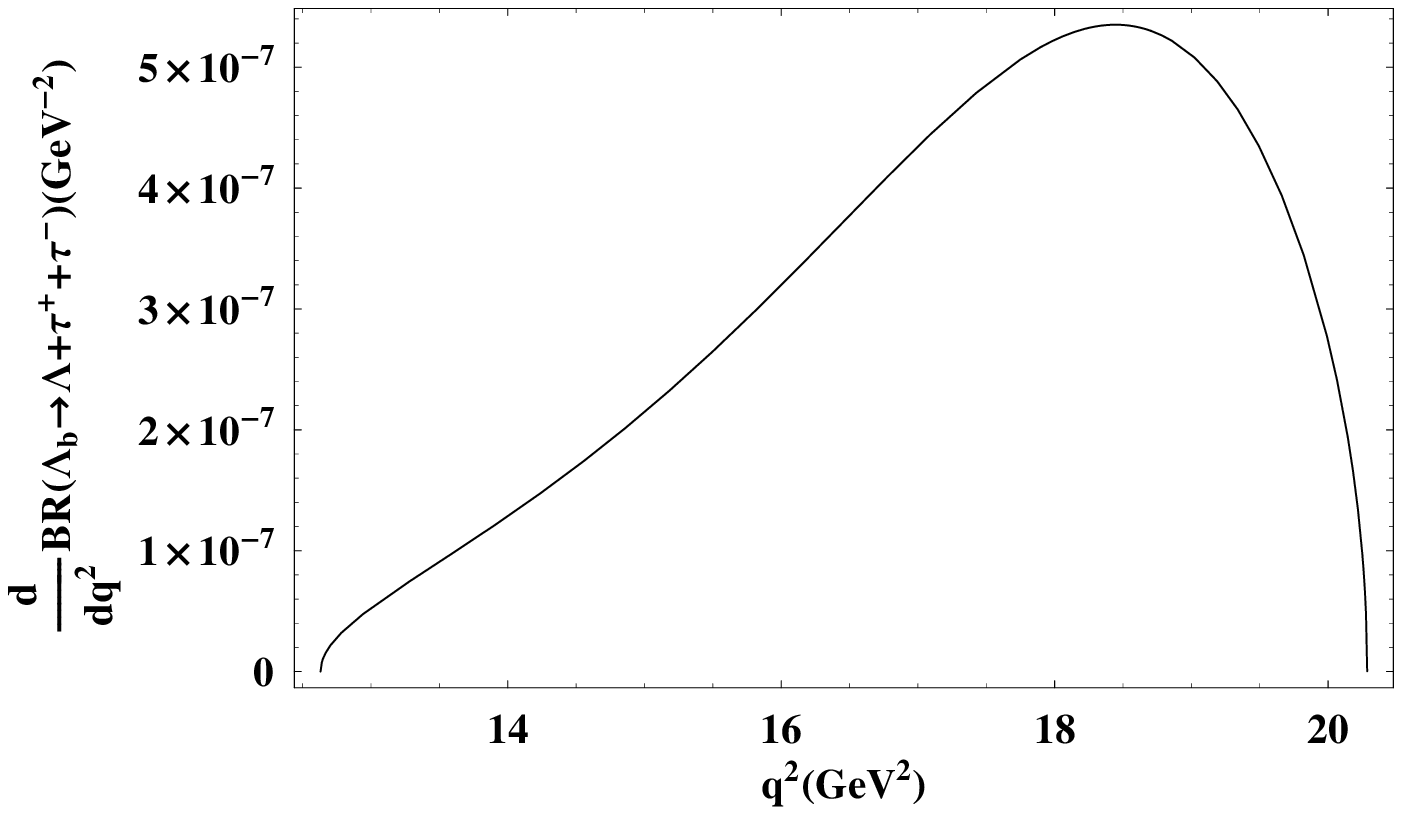}
\put (-350,220){(a)} \put (-100,220){(b)} \put (-350,30){(c)}
\put(-100,30){(d)} \vspace{-1cm}
\end{tabular}
\caption{The differential width for the $\Lambda_b \to \Lambda
l^+l^-$ ($l=\mu, \tau$) decays as functions of $q^2$ with
long-distance   contributions (a, b) and without long-distance
contributions (c, d)  }\label{invariant mass distribution}
\end{center}
\end{figure}

We achieve the invariant dilepton mass distribution for $\Lambda_b
\to \Lambda + l^{+} l^{-}$ ($l=\mu, \tau$) with and without
long-distance contributions as plotted in  Fig. {\ref{invariant mass
distribution}}, where the appearance of large enhancement at the end
point  $q^2=0$  for $\Lambda_b \to \Lambda + \mu^{+} \mu^{-}$ is due
to the factor $1/q^2$ involved in the operator $O_{7 \gamma}$ and
$h(z,s')$ term in Wilson coefficient $C_9^{eff}$in the effective
Hamiltonian for $b \to s l^{+} l^{-}$ transition. The dilepton mass
distribution of $\Lambda_b \to \Lambda + l^{+} l^{-}$ ($l=\mu,
\tau$) without long-distance contributions peaks at higher invariant
masses in Fig. {\ref{invariant mass distribution}}, which is much
different from that given in \cite{c.q. geng 4}. The reason is that
the form factors grow more drastically with the increase of squared
momentum transfer $q^2$ in LCSR than that obtained in QCDSR,   which
is also found  in  the semi-leptonic decay $\Lambda_b \to p l
\bar{\nu}$ \cite{m.q. huang lambda_b to proton}.

Integrating Eq. (\ref{differential decay width}),   we   get the
branching fractions of $\Lambda_b \to \Lambda + l^{+} l^{-}$
($l=\mu, \tau$)   as displayed in Table~\ref{results of
semi-leptonic decay rate}, together with the results  computed in
the framework of QCDSR \cite{c.q. geng 2} and PM  \cite{c.q. geng 2}
for a comparison.
 The QCDSR result
\cite{c.q. geng 2} for decay rate of $\Lambda_b \to \Lambda +
\tau^{+} \tau^{-}$ without long-distance effects  is much smaller
than that presented in this work, which can be attributed to the
larger transition form factors predicted in LCSR due to the
sensitive dependence on the momentum transfer $q^2$.

\begin{table}[h]
\caption{Decay branching ratios ($10^{-6}$ ) of  $\Lambda_b \to
\Lambda + l^{+} l^{-}$ ($l=\mu, \tau$) with and without
long-distance contributions based on the form factors calculated in
this work using light-cone sum rules  and that from QCD sum rules
and pole model}
\begin{center}
\begin{tabular}{c|cc|cc}
  \hline
  \hline
 Model &    \multicolumn{2}{c|}{$\Lambda_b \to \Lambda + \mu^+
  \mu^-$ } &\multicolumn{2}{c}{ $\Lambda_b \to \Lambda + \tau^{+}
  \tau^{-}$} \\
 & without LD &  with LD & without LD &  with LD \\
   \hline LCSR (this work)&  $6.1^{+5.8}_{-1.7}  $ & $39^{+23}_{-11}  $
  & $2.1^{+2.3}_{-0.6}  $ & $4.0^{+3.7}_{-1.1}  $ \\
  \hline
 QCDSR \cite{c.q. geng 2}&   $2.1 $ & $53  $
 & $0.18  $ & $11  $ \\
  \hline
  PM \cite{c.q. geng 2}&   $1.2  $ & $36 $
  & $0.26 $
  &   $9.0  $ \\
 \hline
  \hline
\end{tabular}
\end{center}
\label{results of semi-leptonic decay rate}
\end{table}

 We are now ready to investigate the effects of  magnetic
penguin operator $O_{7 \gamma}$ in the semi-leptonic decay
$\Lambda_b \to \Lambda + l^{+} l^{-}$ ($l=\mu, \tau$). Without the
operator $O_{7 \gamma}$, the $BR(\Lambda_b \to \Lambda l^{+} l^{-})$
is 16 \% and 27 \% smaller for final states being $\mu^{+}\mu^{-}$
and $\tau^{+}\tau^{-}$, respectively. However, the branching
fractions can be 38 \% and 57 \% smaller for these two modes,
respectively, if we adopt the same magnitude of $C_7^{eff}$ as that
in the SM but with an opposite sign. This   confirms the conclusion
  of ref. \cite{c.q. geng 4}   that
$BR(\Lambda_b \to \Lambda l^{+} l^{-})$ can serve as  a promising
quantity to explore the new physics effects as well as constrain
parameter space of various models beyond the SM.

  Roughly speaking, the long-distance effects
on the decay rate for semi-leptonic decays of $\Lambda_b \to \Lambda
+ l^{+} l^{-}$ can be given by
\begin{equation}
BR_{LD}(\Lambda_b \to \Lambda + l^{+} l^{-})=\sum_i BR(\Lambda_b \to
\Lambda +\Psi_{i}) \times  BR(\Psi_{i} \to l^{+} l^{-}) \,\, .
\end{equation}
Utilizing the experimental data on the leptonic decays  of $J/\psi$
and $\Psi(2S)$
 \cite{PDG},
\begin{equation}
\begin{array}{ll}
BR(J/\Psi \to \mu^{+} \mu^{-})= (5.93 \pm 0.06) \% \, , &
BR(\Psi(2S) \to \mu^{+} \mu^{-})= (7.3 \pm 0.8) \times 10^{-3} \, , \nonumber \\
BR(\Psi(2S) \to \tau^{+} \tau^{-})= (2.8 \pm 0.7) \times 10^{-3} \,,
\end{array}
\end{equation}
we can achieve the following relation
\begin{eqnarray}
{BR_{LD}(\Lambda_b \to \Lambda + \mu^{+} \mu^{-})  \over
BR_{LD}(\Lambda_b \to \Lambda + \tau^{+} \tau^{-})}&=&{ BR(J/\Psi
\to \mu^{+} \mu^{-}) +BR(\Psi(2S) \to \mu^{+} \mu^{-}) \over
BR(\Psi(2S) \to \tau^{+} \tau^{-})} \nonumber
\\ & = & 23.8 ,
\end{eqnarray}
where the assumption $BR(\Lambda_b \to \Lambda + J/\psi) =
BR(\Lambda_b \to \Lambda + \psi (2S) )$ has been used in the above
derivations. We can see that this naive estimation is consistent
with that computed in this work, but   the number got in Ref.
\cite{c.q. geng 2} is somewhat smaller.

  We then concentrate on the role of  long-distance effects played
in the  decay rates for both final states    $\mu^{+} \mu^{-}$ and
$\tau^{+} \tau^{-}$ within LCSR approach. As  can be seen,
$\Lambda_b \to \Lambda + \mu^{+} \mu^{-}$ decay is dominated by the
long-distance contributions in the vicinity of the $c \bar{c}$
resonance region, which is approximately $6 \sim 7$ times larger
than that from short-distance contributions. In contrast to the case
of final states with $\mu^{+} \mu^{-}$, the long-distance
contributions are almost the same size as that from short-distance
in $\Lambda_b \to \Lambda + \tau^{+} \tau^{-}$ decay, which differs
  from that  in the other two methods.

\subsection{$\Lambda$ polarization asymmetry of $\Lambda_b \to \Lambda + l^{+} l^{-}$}

Similar to the radiative decay $\Lambda_b \to \Lambda + \gamma$, we
can write the differential decay width  for $\Lambda_b \to \Lambda +
l^{+} l^{-}$ with respect to the squared momentum transfer $q^2$ in
the following form
\begin{equation}
{d\Gamma(\Lambda_b \to \Lambda l^{+} l^{-}) \over dq^2} =A(q^2)+
{p_{\Lambda_b} \cdot s} B(q^2).
\end{equation}
The  $q^2$ dependence of the function ``${{\bf p}_{\Lambda} \over
E_{\Lambda}} B(q^2)$" describing the polarization of $\Lambda$
baryon is plotted in Fig. \ref{spin dependent_function} for both the
cases with and without LD contributions. The function $A(q^2)$ is
just the invariant dilepton mass distributions discussed in the
previous subsection.

\begin{figure}
\begin{center}
\begin{tabular}{ccc}
\vspace{-2cm}
\includegraphics[scale=0.6]{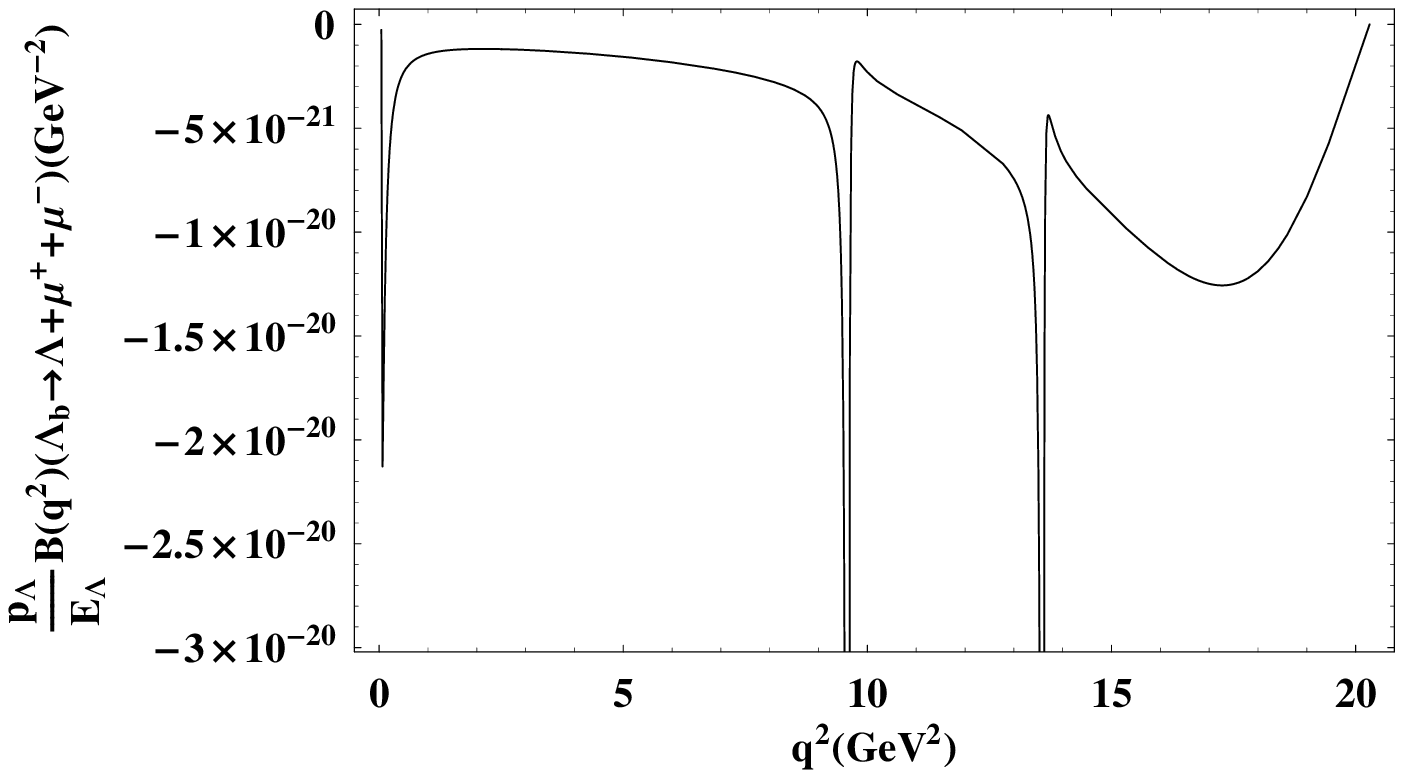}
\includegraphics[scale=0.6]{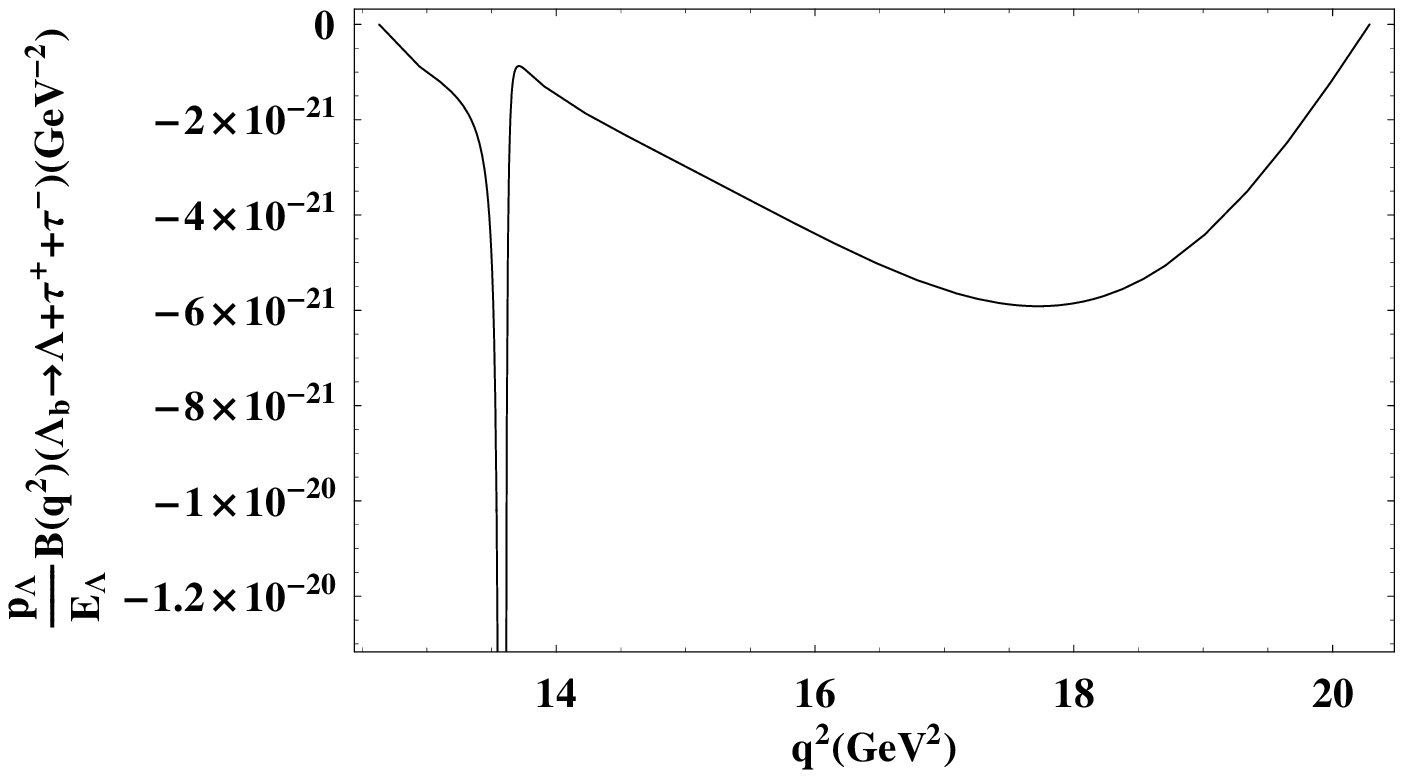}
\\
\includegraphics[scale=0.6]{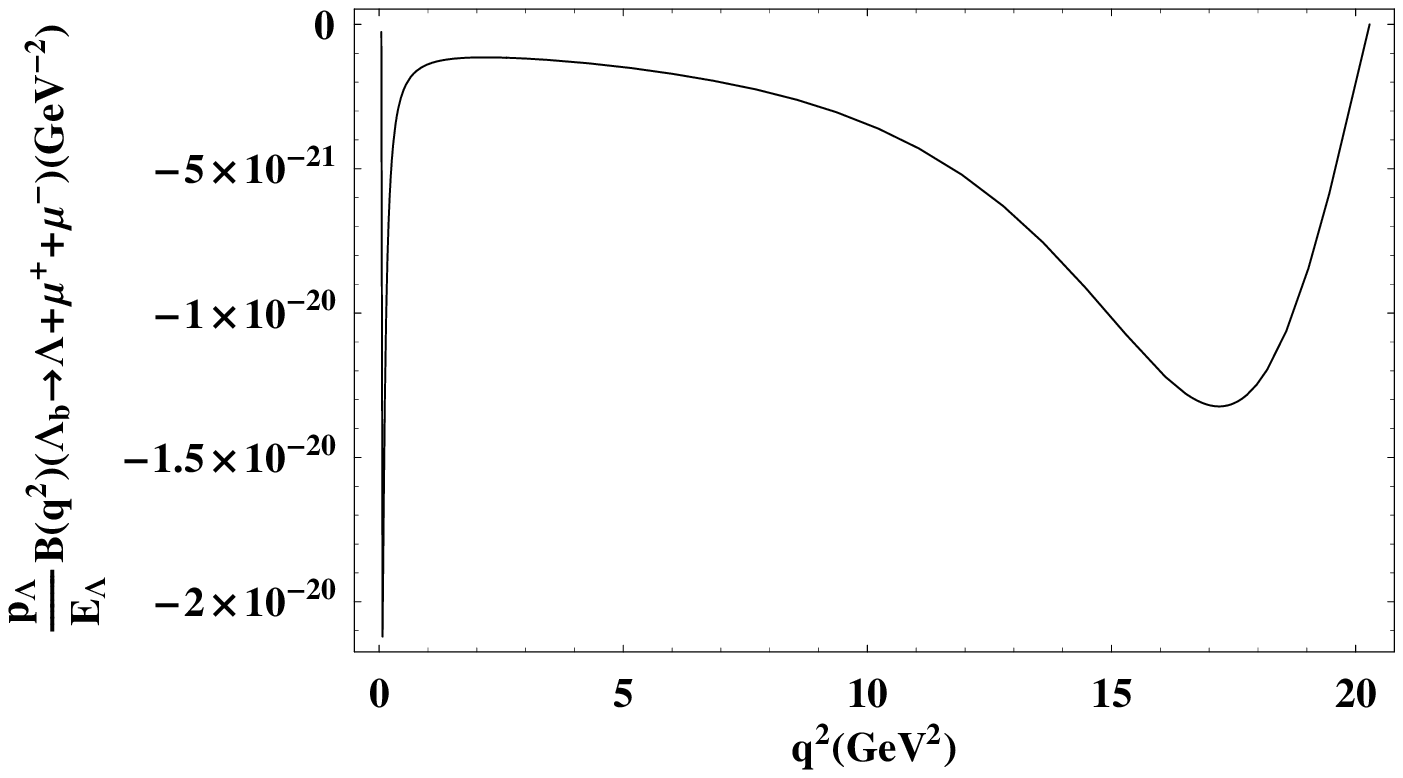}
\includegraphics[scale=0.6]{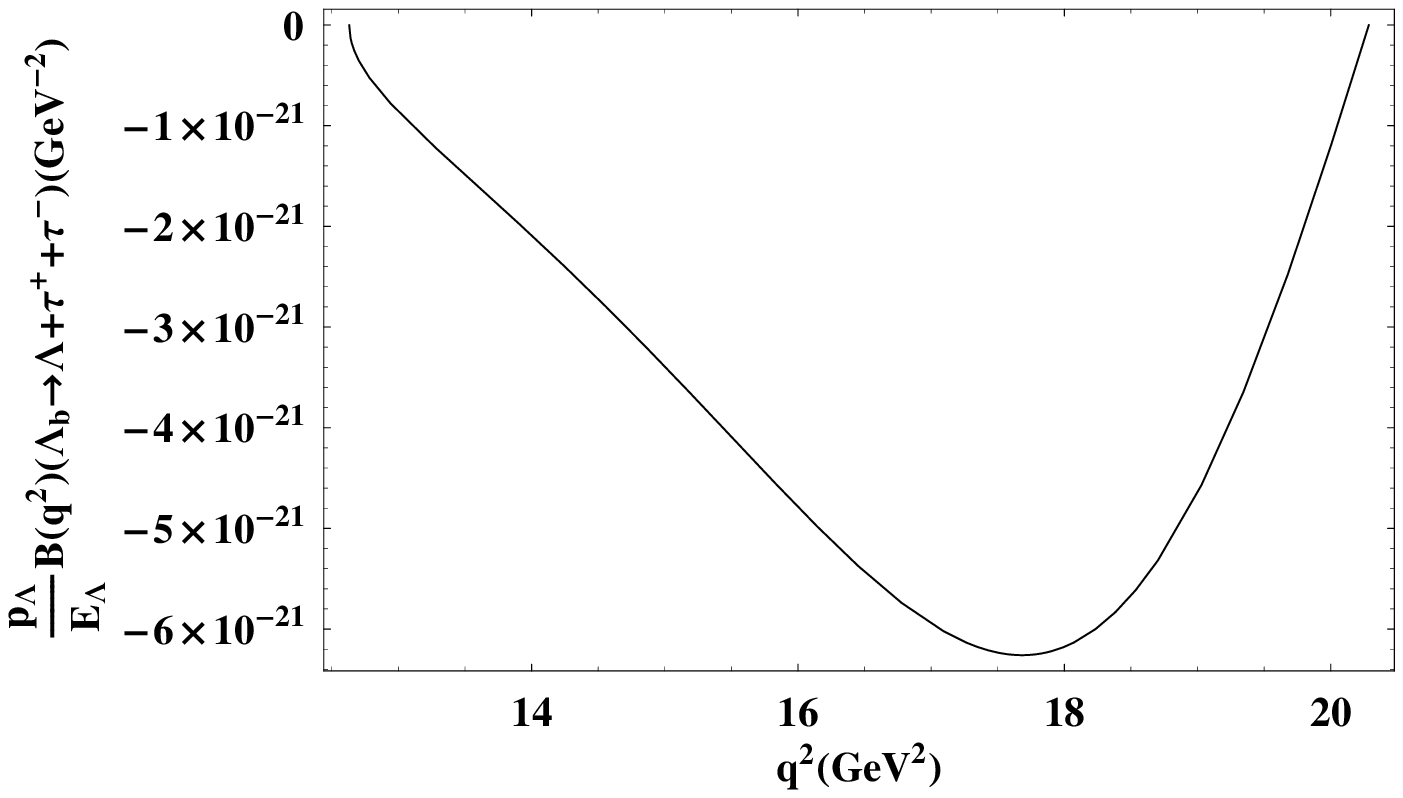}
\put (-350,220){(a)} \put (-100,220){(b)} \put (-350,30){(c)}
\put(-100,30){(d)} \vspace{-1cm}
\end{tabular}
\caption{Spin-dependent term ${{\bf p}_{\Lambda} \over E_{\Lambda}}
B(q^2)$ as a function of $q^2$ with long-distance contributions (a,
b) and without long-distance contributions  (c, d)   }\label{spin
dependent_function}
\end{center}
\end{figure}

Integrating the differential decay width over $q^2$, the  decay
width of $\Lambda_b \to \Lambda + l^{+} l^{-}$ with polarized
$\Lambda$ baryon can be written in the form of Eq. (\ref{form of
polarization asymmetry})
\begin{eqnarray}
\Gamma(\Lambda_b \to \Lambda l^{+} l^{-}) = {1 \over 2}\Gamma_0
[1+\alpha \,\, {\bf{\hat{p}}} \cdot {\bf{s}} ]={1 \over 2}\Gamma_0
[1+\alpha^{\prime} \,\, {\bf{\hat{p}}} \cdot {\bf{\hat{\xi}}} ],
\end{eqnarray}
where $\Gamma_0$ is the total decay width of $\Lambda_b \to \Lambda
+ l^{+} l^{-}$
\begin{eqnarray}
\Gamma_0=2 \int_{q^2_{min}}^{{q^2_{max}}} A(q^2) d q^2 ,
\end{eqnarray}
with $q^2_{min}= 4 m_l^2$ and $q^2_{max}=
(m_{\Lambda_b}-m_{\Lambda})^2$. Besides, $\alpha$ and
$\alpha^{\prime}$ are the polarization asymmetry parameters, whose
manifest expressions can be expressed in terms of function  $B(q^2)$
as
\begin{eqnarray}
\alpha= \frac{2m_{\Lambda_b}}{\Gamma_0}
{\int_{q^2_{min}}^{q^2_{max}} {|{\bf p}_{\Lambda}| \over
E_{\Lambda}} B(q^2) dq^2 }, \qquad \alpha^{\prime}=
\frac{2m_{\Lambda_b} }{\Gamma_0}{\int_{q^2_{min}}^{q^2_{max}} {|{\bf
p}_{\Lambda}| \over m_{\Lambda}} B(q^2) dq^2},
\end{eqnarray}
with $|{\bf p}_{\Lambda}| =
\sqrt{({m_{\Lambda_b}^2+m_{\Lambda}^2-q^2 \over 2 m_{\Lambda_b}})^2
- m_{\Lambda}^2}$ being the magnitude of three-momentum for
$\Lambda$ baryon in the rest frame of $\Lambda_b$. The numerical
results of polarization variables $\alpha$ and $\alpha^{\prime}$ are
grouped in Table~\ref{results of polarization asymmetry}. As can be
seen, the long-distance contributions   have small effects on the
polarization asymmetry for decay of $\Lambda_b \to \Lambda +
\tau^{+} \tau^{-}$, however, they can modify the polarization
asymmetry for the mode $\Lambda_b \to \Lambda + \mu^{+} \mu^{-}$
remarkably. The reason is that only $c \bar{c}$ resonances above
(including) the $\Psi(2S)$ threshold can have influences on the
long-distance contributions for $\Lambda_b \to \Lambda + \tau^{+}
\tau^{-}$, which are much suppressed compared with that from
$J/\psi$ in the decay of $\Lambda_b \to \Lambda + \mu^{+} \mu^{-}$.
 The value of $\alpha$ corresponding to the channel $\Lambda_b \to
\Lambda + \mu^{+} \mu^{-}$ is consistent with
$-0.54^{+0.04}_{-0.04}$ that calculated in \cite{Huang} within the
error bar.

\begin{table}
\caption{Polarization asymmetry parameter $\alpha$ and
$\alpha^{\prime}$ of $\Lambda$ baryon with and without long-distance
contributions based on the form factors calculated in light-cone sum
rules }
\begin{center}
\begin{tabular}{c|cc|cc}
  \hline
  \hline
  & \multicolumn{2}{c|}{$\Lambda_b \to \Lambda + \mu^{+}
  \mu^{-}$ }& \multicolumn{2}{c}{$\Lambda_b \to \Lambda + \tau^{+}
  \tau^{-}$} \\
&  without LD & with LD &without LD & with LD\\
   \hline
  $\alpha$ & $-0.36^{+0.05}_{-0.02}$ & $-0.50^{+0.04}_{-0.01}$
   & $-0.28^{+0.03}_{-0.03}$ & $-0.27^{+0.03}_{-0.03}$\\
   $\alpha^{\prime}$         & $-0.55^{+0.09}_{-0.04}$  & $-0.88^{+0.07}_{-0.03}$
    & $-0.36^{+0.04}_{-0.03}$   & $-0.39^{+0.06}_{-0.03}$ \\
  \hline
  \hline
\end{tabular}
\end{center}
\label{results of polarization asymmetry}
\end{table}

\subsection{Forward-backward asymmetry of $\Lambda_b \to \Lambda + l^{+} l^{-}$}

\begin{figure}[htbp]
   \begin{center}\vspace{-8cm}
\hspace{0cm}     \epsfig{file=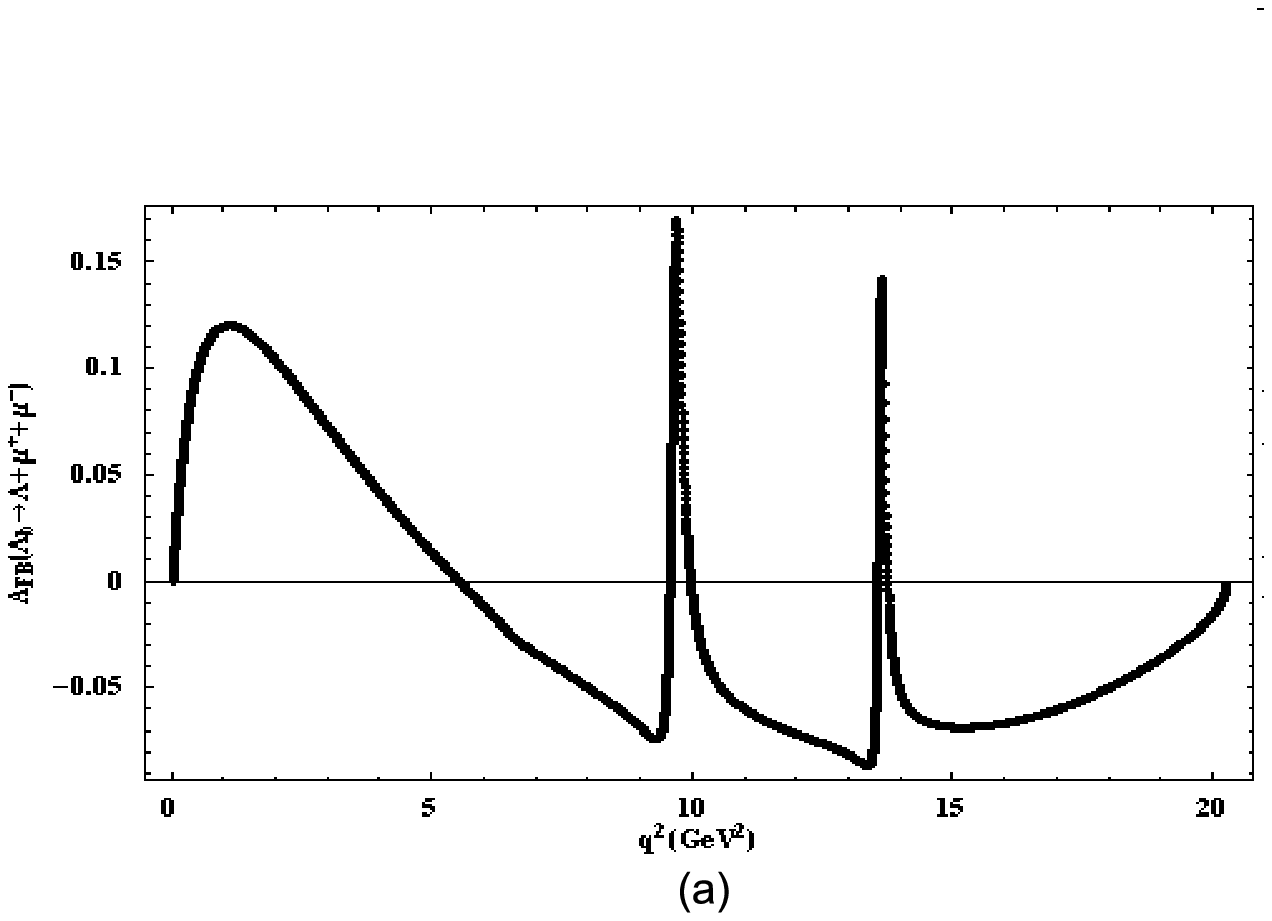, bbllx=5.8cm,bblly=10.4cm,bburx=11cm,bbury=23cm,%
 width=3.5cm,angle=0}\\
{\hspace{2cm}
\epsfig{file=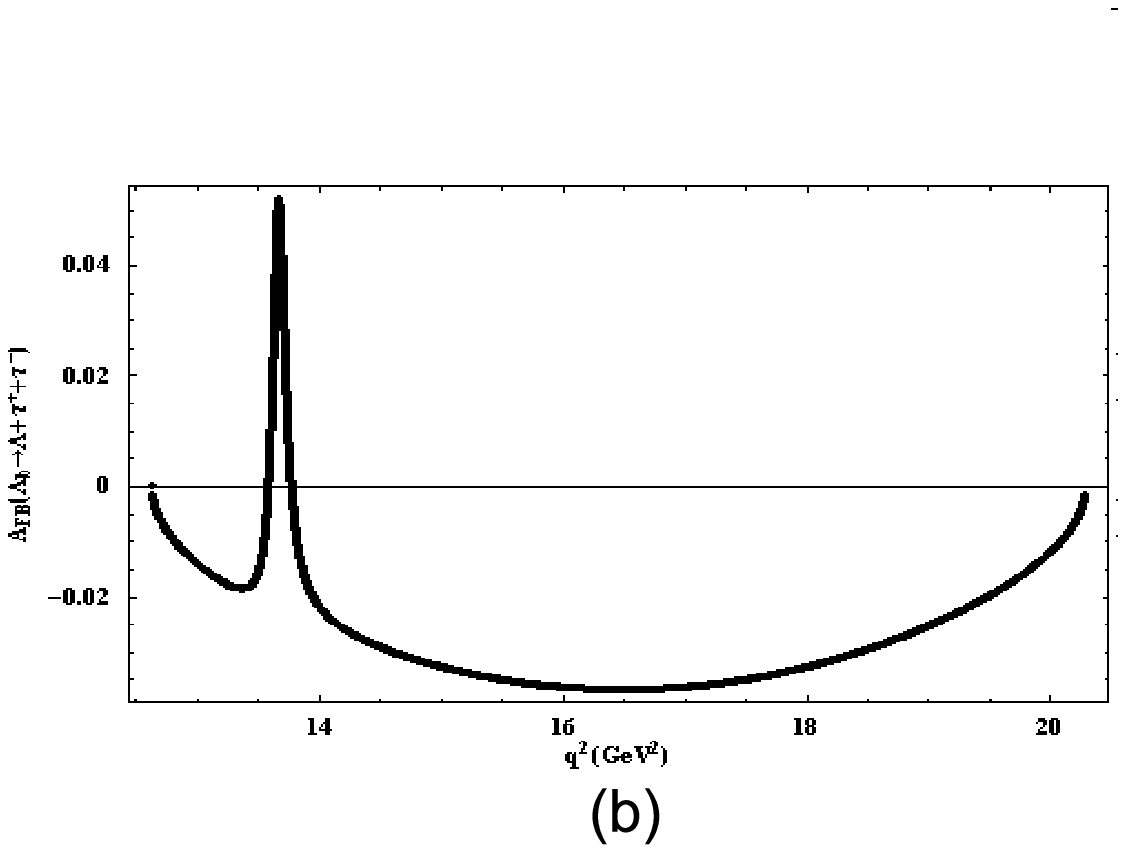, bbllx=5.8cm,bblly=10.4cm,bburx=11cm,bbury=23cm,%
 width=4cm,angle=0}}
   \end{center}
  \vspace{8cm} \caption{Normalized forward-backward asymmetry $A_{FB}(q^2)$ as a
function of $q^2$ with long-distance contributions: (a) for muon,
and (b) for tauon.}
   \label{normalized_FBA1}
  \end{figure}

\begin{figure}[htbp]
   \begin{center}\vspace{-8cm}
\hspace{2cm}     \epsfig{file=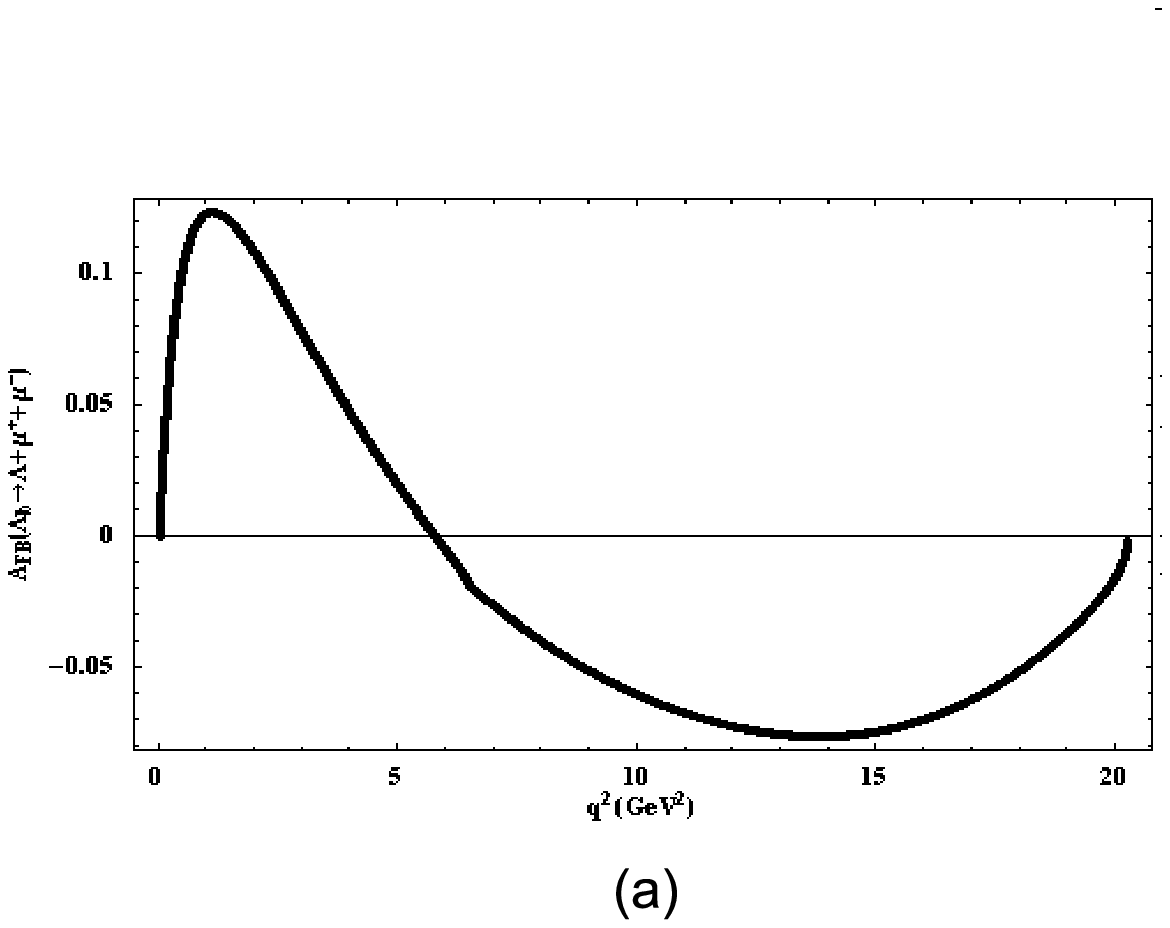, bbllx=5.8cm,bblly=10.4cm,bburx=11cm,bbury=23cm,%
 width=4.0cm,angle=0}\\
{\hspace{2cm}
\epsfig{file=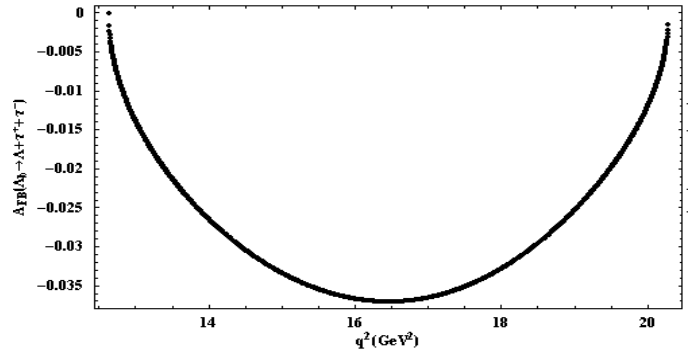, bbllx=5.8cm,bblly=10.4cm,bburx=11cm,bbury=23cm,%
 width=4cm,angle=0}}
   \end{center}
  \vspace{8cm} \caption{Normalized forward-backward asymmetry $A_{FB}(q^2)$ as a
function of $q^2$ without long-distance contributions: (a) for muon,
and (b) for tauon.}
   \label{normalized-FBA2}
  \end{figure}

For the illustration of forward-backward asymmetry, we consider the
following  double patrial differential decay rates for the decays of
$\Lambda_b \to \Lambda l^{+} l^{-}$
\begin{eqnarray}
{d^2 \Gamma (q^2, z) \over dq^2 dz }={1 \over (2 \pi)^3} {1 \over 64
m_{\Lambda_b}^3} \lambda^{1/2}(m_{\Lambda_b}^2,m_{\Lambda}^2, q^2)
\sqrt{1-{4 m_l^2 \over q^2}} |{\widetilde{M}}_{\Lambda_b \to \Lambda
l^+ l^-}|^2\,\,,
\end{eqnarray}
where $z={\rm cos} \theta$ and $\theta$ is the angle between the
momentum of $\Lambda_b$ baryon and $l^{-}$ in the dilepton rest
frame; $\lambda(a,b,c)=a^2+b^2+c^2-2ab-2ac-2bc$. Following Refs.
\cite{c.q. geng 4, b to s in theory 9}, the differential and
normalized forward-backward asymmetries for the semi-leptonic decay
$\Lambda_b \to \Lambda l^{+} l^{-}$ can be defined as
\begin{eqnarray}
{d A_{FB}(q^2) \over d q^2}=\int_0^1 dz  {d^2 \Gamma (q^2, z) \over
dq^2 dz} - \int_{-1}^0 dz  {d^2 \Gamma (q^2, z) \over dq^2 dz}.
\end{eqnarray}
and
\begin{eqnarray}
A_{FB}(q^2)={ \int_0^1 dz  {d^2 \Gamma (q^2, z) \over dq^2 dz} -
\int_{-1}^0 dz  {d^2 \Gamma (q^2, z) \over dq^2 dz}\over \int_0^1 dz
{d^2 \Gamma (q^2, z) \over dq^2 dz} + \int_{-1}^0 dz  {d^2 \Gamma
(q^2, z) \over dq^2 dz} }.
\end{eqnarray}
Making use of the decay amplitude in Eq. (\ref{b to s l l}), the
differential forward-backward asymmetry for decays of $\Lambda_b \to
\Lambda + l^{+} l^{-}$ can be calculated as

\begin{eqnarray}
{d A_{FB}(q^2) \over d q^2}={ G_F^2 \alpha_{em}^2 |V_{tb}
V_{ts}^{\ast}|^2 \over 256 m_{\Lambda_b}^3 \pi^5}
\lambda(m_{\Lambda_b^2},m_{\Lambda}^2,q^2) (1-{4 m_l^2 \over q^2})
R_{FB}(q^2),
\end{eqnarray}
with
\begin{eqnarray}
R_{FB}(q^2)&=&2[(m_s m_{\Lambda}+m_b m_{\Lambda_b}) f_2^2- m_s
(m_{\Lambda}^2-m_{\Lambda_b}^2+q^2) f_2 g_2  + (m_s m_{\Lambda}-m_b
m_{\Lambda_b}) q^2 g_2^2] {\rm {Re}}(C_7^{eff} C_{10}^{\ast}) \nonumber \\
&&+[(f_2 -g_2 m_{\Lambda})^2-g_2^2 m_{\Lambda_b}^2]q^2 {\rm
{Re}}(C_9^{eff} C_{10}^{\ast}), \label{FBA expressions}
\end{eqnarray}
where we have retained masses for both the lepton and strange quark.
In the limit of $m_s \to 0$, our results will be  the same as that
in Ref. \cite{c.q. geng 4}. The differential forward-backward
asymmetry for semi-leptonic decay of  $\Lambda_b \to \Lambda l^{+}
l^{-}$ only depends on the following two combinations of Wilson
coefficients ${\rm {Re}}(C_7^{eff} C_{10}^{\ast})$ and ${\rm
{Re}}(C_9^{eff} C_{10}^{\ast})$, since only terms involving ${\rm
{Tr}}[L_{\mu}^{V}L_{\nu}^{A}]$ in the differential decay width can
give rise to one power of ``${\rm{cos}} \theta$" \cite{c.q. geng 4},
where $L_{\mu}^{V}$ and $L_{\nu}^{A}$ represent the vector and
axial-vector currents of leptonic sector. The zero position $t_0$ of
forward-backward asymmetry is determined by the relation
\begin{eqnarray}
{\rm {Re}}(C_9^{eff} C_{10}^{\ast})= -{ 2 m_b m_{\Lambda_b} \over
t_0} { f_2^2 -t_0 g_2^2 \over (f_2-m_{\Lambda}g_2)^2-m_{\Lambda_b}^2
g_2^2} {\rm {Re}}(C_7^{eff} C_{10}^{\ast}),
\end{eqnarray}
where the small number of strange quark mass is neglected. It is
easy to observe that $t_0$  only rely on the  Wilson coefficient as
well as the ratio of form factors. Utilizing the form factors
computed in this work, we can derive the zero position of
forward-backward asymmetry at $t_0= 6.0 {\rm{GeV^2}}$ without
long-distance contributions, which indicates that the
forward-backward asymmetry for $\Lambda_b \to \Lambda + \tau^{+}
\tau^{-}$ could not be zero apart from the resonance regions and end
points. However, the number of $t_0$ can shift to $3.5 {\rm{GeV^2}}$
using the form factors calculated in QCDSR \cite{Huang}, due to the
quite different predictions on the ratio of from factors in these
two approaches. It is mentioned in \cite{c.q. geng 4,Hiller:2001zj}
that the zero-position of forward-backward asymmetry is not
sensitive to the form factors in the large energy limit. However,
power corrections contributed by the higher twist distribution
amplitudes of $\Lambda$ baryon can bring about significant effects
on the ratio of form factors, which can even overwhelm the  sign of
form factor breaking the heavy quark symmetry. The position of the
zero of the  forward-backward  can shift observably  for the ratio
of form factors with different sign. The distribution of normalized
forward-backward asymmetry as a function of $q^2$ with and without
long-distance contributions is presented in Fig.
\ref{normalized_FBA1} and Fig. \ref{normalized-FBA2}.

It is also useful to introduce the integrated forward-backward
asymmetry $\langle A_{FB} \rangle$  in order to characterize the
typical value of forward-backward asymmetry \cite{c.q. geng 4}
\begin{eqnarray}
\langle A_{FB} \rangle= \int_{t_{min}}^{t_{max}}A_{FB}(t) dt ,
\end{eqnarray}
with $t=q^2/{m_{\Lambda_b}^2}$, $t_{min}={4 m_l^2 /m_{\Lambda_b}^2}$
and  $t_{max}={(m_{\Lambda_b}^2-m_{\Lambda}^2) /m_{\Lambda_b}^2}$.
The numerical results of integrated forward-backward asymmetry are
grouped in Table \ref{results of averaged FBA}, where the
evaluations in QCDSR  are also given for a comparison. As can be
observed, the predictions on the $\langle A_{FB} \rangle$ in LCSR
are typically one order smaller than that in QCDSR due to the quite
different predictions on the ratio of form factors. The large
discrepancy  may be recoiled by including the higher conformal spin
contributions in the sum rules of form factors on the light-cone as
well as radiative corrections  for both two types of sum rules.

\begin{table}[h]
\caption{Averaged forward-backward asymmetry (\%) for $\Lambda_b \to
\Lambda + l^{+} l^{-}$ with and without long-distance contributions
based on the form factors calculated in this work using light-cone
sum rules  and that from QCD sum rules}
\begin{center}
\begin{tabular}{c|ccc}
  \hline
   \hline
  model &    & $\Lambda_b \to \Lambda + \mu^{+}
  \mu^{-}$ & $\Lambda_b \to \Lambda + \tau^{+}
  \tau^{-}$ \\
  \hline
LCSR  & without LD  & $-1.22^{+1.42}_{-0.73}  $  & $-0.67^{+0.23}_{-0.21}  $  \\
(this work)  & with LD & $-0.99^{+1.32}_{-0.68}  $ & $-0.62^{+0.22}_{-0.21}  $ \\
\hline
QCDSR \cite{c.q. geng 4} & without LD  & $-14.38^{+1.14}_{-0.0}   $
& $-3.99^{+0.01}_{-0.01}  $  \\
  \hline
   \hline
\end{tabular}
\end{center}
\label{results of averaged FBA}
\end{table}

\section{Summary}

The   study on   rare decay of $\Lambda_b \to \Lambda + \gamma$ and
$\Lambda_b \to \Lambda + l^{+} l^{-}$ can serve as the baryonic
counterparts of analysis on $B \to K^{\ast} \gamma$ \cite{b to s in
theory 31,b to s in theory 32,b to s in theory 33,b to s in theory
34,b to s in theory 35,b to s in theory 36,b to s in theory 37,b to
s in theory 38,b to s in theory 39,b to s in theory 40,b to s in
theory 41}, $B \to K^{\ast}  l^{+} l^{-}$ \cite{b to s in theory
16,b to s in theory 17,b to s in theory 18,b to s in theory 19,b to
s in theory 20,b to s in theory 21} and could also be   the
extension of investigations on heavy baryon decays $\Lambda_b \to p
l \bar{ \nu}$ \cite{m.q. huang lambda_b to proton} and $\Lambda_c
\to \Lambda l \bar{ \nu}$ \cite{m.q. huang}. Such decays play the
role as a corner stone \cite{neubert} to explore the quark-flavor
structure of the SM as well as determine  its fundamental parameters
such as the CKM matrix.

Although we have achieved inspiring progress in the heavy meson
decays with the help of heavy quark expansions and factorization
techniques, serious and systemical studies on heavy baryon decays
are comparatively behind on account of the complexity of inner
structures for baryon systems. In particular, the predictions on the
physical observables associating with heavy baryon decays can vary
even by the orders of magnitude employing different theoretical
tools.  In this work, we explore the soft contributions to the form
factors responsible for the $\Lambda_b \to \Lambda$ transition in
terms of LCSR approach  to understand the tremendous discrepancy on
predictions of decay rate for $\Lambda_b \to \Lambda + \gamma$
between   theoretical methods. More importantly, power corrections
from the higher twist distribution amplitudes of $\Lambda$ baryon to
the transition form factors are also investigated to the leading
conformal spin in detail. It is observed that the higher twist LCDAs
almost have no influences on the transition form factors reserving
the heavy quark spin symmetry, while such corrections can result in
significant impacts on the from factors breaking the heavy quark
spin symmetry. In addition, we also find that various models for the
distribution amplitudes of $\Lambda$ baryon  can lead to quite
different predictions on the transition form factors, which will be
tested  by the  experiments on Tevatron and LHC     in the future.

 We   confirm the conclusion
of Ref. \cite{x.g. he,Huang} that the $\Lambda$ polarization
asymmetry of $\Lambda_b \to \Lambda \gamma$ only relies on the
relative strength of left- and right- handed couplings between
quarks and is free of the pollution from the strong interaction. We
also discuss some interesting observables in phenomenology, such as
decay rate, dilepton distribution, polarization asymmetry and
forward-backward asymmetry of $\Lambda_b \to \Lambda l^{+} l^{-}$.
We use the pole model  to extrapolate the results of form factors
calculated in LCSR to the whole physical region in view of the
failure of light-cone expansion for correlation function in the
large  momentum transfer region. Our results indicate that the decay
rate for $\Lambda_b \to \Lambda \mu^{+} \mu^{-}$ is about five times
larger than that for $\Lambda_b \to \Lambda \gamma$ due to the
long-distance contributions from the charmonium  resonance region.
The polarization asymmetry of $\Lambda_b \to \Lambda \mu^{+}
\mu^{-}$ is much sensitive to the long-distance contributions than
that of $\Lambda_b \to \Lambda \tau^{+} \tau^{-}$, since leading
resonance $J/\psi$ does not contribute to the case of final state
with tauon pair. As for the integrated forward-backward asymmetry,
our results for the magnitude of  both  muon and tauon cases are
typically one order smaller than that given in QCDSR due to the
quite different predictions on the ratio of transition form factors,
which can also shift the zero position of forward-backward asymmetry
from $t_0= 6.0 {\rm{GeV^2}}$ in LCSR to $3.5 {\rm{GeV^2}}$ predicted
by QCDSR approach.


\section*{Acknowledgements}
This work is partly supported by National Science Foundation of
China under Grant No.~10735080 and  10625525. The authors would like
to thank T.M. Aliev, C.H. Chen, H.Y. Cheng, X.G. He, G. Hiller, C.S.
Huang, M.Q. Huang, T. Huang, M. Jamil, A. Lenz and Y.L. Liu for
helpful discussions.

\appendix

\section{QCD sum rules for $f_{\Lambda}$, $f_{\Lambda_b}$ and $\lambda_1$}
\label{decay constants of lambda}

The sum rules of $f_{\Lambda}$ and $\lambda_1$ have been derived in
\cite{m.q. huang}, we would like to collect them in this appendix
for the completeness of the paper.
\begin{eqnarray}
(4\pi)^4f_\Lambda^2e^{-M^2/M_B^2}&=&\frac{2}{5}\int_{{m_s^2}}^{s_\Lambda^0}
s(1-x)^5e^{-s/M_B^2}ds-\frac{b}{3}\int_{{m_s^2}}^{s_\Lambda^0}
x(1-x)(1-2x)e^{-s/M_B^2}\frac{ds}{s}, \label{f_lambda}
\end{eqnarray}
\begin{eqnarray}
4(2\pi)^4\lambda_1^2M^2e^{-M^2/M_B^2}&=&\frac{1}{2}\int_{{m_s^2}}^{s_0}
s^2\left[(1-x^2)(1-8x+x^2)-12x^2\ln\;x\right]e^{-s/M_B^2}ds\nonumber\\
&+&\frac{b}{4}\int_{{m_s^2}}^{s_0}
(1-x)^2e^{-s/M_B^2}ds-\frac{4}{3}\;a^2e^{-M^2/M_B^2},
\end{eqnarray}
where $x=m_s^2/s$ and $m_s$ is the mass of strange quark. The values
of the non-perturbative condensate at scale $\mu=1 {\rm GeV}$ are
given by
\begin{eqnarray}
 a &=& - (2\pi)^2 \langle \bar q q \rangle = 0.55 \;
{\rm GeV^3} \,,
\nonumber \\
b &=& (2\pi)^2 \langle \frac{\alpha_s}{\pi} G^2\rangle = 0.47 \;
{\rm GeV^4} \,.
\end{eqnarray}
The sum rules for $f_{\Lambda_b}$ can be read from Eq.
(\ref{f_lambda}), only with the following replacement:
\begin{eqnarray}
f_{\Lambda} \to f_{\Lambda_b}, \,\,\,  m_s \to m_b, \,\,\,
s_{\Lambda}^0 \to  s_{\Lambda_b}^0.
\end{eqnarray}
With the threshold value $s_{\Lambda}^0=1.6 {\rm GeV}^2$, we can
arrive at the numbers of $f_{\Lambda}$ and  $\lambda_1$ as
\begin{eqnarray}
f_\Lambda=6.0^{+0.4}_{-0.4} \times 10^{-3} {\rm GeV}^2,\,\,\,
\lambda_1=-1.3^{+0.2}_{-0.2} \times 10^{-2} {\rm GeV}^2,
\end{eqnarray}
within the Borel window $M_B^2 \in [1.0, \,\, 2.0] {\rm GeV}^2$. In
the same way, the result of $f_{\Lambda_b}$ can be derived as
$f_{\Lambda_b}=3.9^{+0.4}_{-0.2} \times 10^{-3} {\rm{GeV}}^{2}$ with
the selected threshold parameter $s_{\Lambda_b}^0=39 \pm 1
{\rm{GeV}}^2$ and Borel mass $M_B^2 \in [2.0, \,\, 3.5] {\rm
GeV}^2$.

\end{document}